\documentclass[aps, 10pt, superscriptaddress, notitlepage,longbibliography]{revtex4-1}

\pdfoutput=1
\usepackage{graphicx}
\usepackage{amssymb,amsmath}
\usepackage{bm}

\begin{document}

\title{Sparse optimization of mutual synchronization in collectively oscillating networks}

\author{Hiroya Nakao\footnote{Corresponding author: nakao@sc.e.titech.ac.jp}}
\affiliation{Department of Systems and Control Engineering,
Tokyo Institute of Technology, Tokyo 152-8552, Japan}

\author{Katsunori Yamaguchi}
\affiliation{Department of Systems and Control Engineering,
Tokyo Institute of Technology, Tokyo 152-8552, Japan}

\author{Shingo Katayama}
\affiliation{Department of Systems and Control Engineering,
Tokyo Institute of Technology, Tokyo 152-8552, Japan}

\author{Tatsuo Yanagita}
\affiliation{Department of Engineering Science, Osaka Electro-Communication University, Neyagawa, Osaka 572-8530, Japan}

\date{\today}

\begin{abstract}
We consider a pair of collectively oscillating networks of dynamical elements and optimize their internetwork coupling for efficient mutual synchronization based on the phase reduction theory
developed in {Ref.~[H. Nakao, S. Yasui, M. Ota, K. Arai, and Y. Kawamura, Chaos {\bf 28}, 045103 (2018)].}
The dynamical equations describing a pair of weakly coupled networks are reduced to a pair of coupled phase equations, and the linear stability of the synchronized state between the networks is represented as a function of the internetwork coupling matrix.
We seek the optimal coupling by minimizing the {Frobenius} and $L_1$ norms of the internetwork coupling matrix for the prescribed linear stability of the synchronized state.
Depending on the norm, either a dense or sparse internetwork coupling yielding efficient mutual synchronization of the networks is obtained.
In particular, a sparse yet resilient internetwork coupling is obtained by $L_1$-norm optimization with additional constraints on the individual connection weights.
\end{abstract}

\maketitle

{\sf Synchronization of collective oscillations in networks of dynamical elements is often important for the functioning of various real-world systems. In this study, we consider mutual synchronization between a pair of networks exhibiting stable collective oscillations and optimize their internetwork coupling based on phase reduction theory. We derive an optimal internetwork coupling that achieves the prescribed linear stability of in-phase synchronization under two different criteria. In particular, we demonstrate that sparse yet resilient coupling can be obtained by $L_1$-norm optimization with additional constraints on the connection weights. The results obtained can benefit the design of coupled dynamical networks with efficient synchronization properties.}

\section{Introduction}

Various networks of coupled dynamical elements exhibiting collective oscillations exist in the real world, e.g., gene-regulatory networks~\cite{goldbeter1997biochemical}, neuronal networks~\cite{buzsaki2006rhythms}, and power grids~\cite{motter2013spontaneous}. Even a simple electronic circuit that generates periodic signals may be considered a collectively oscillating network, where individual electronic components are the network nodes.
The synchronization of collective oscillations in such networks of dynamical elements is often important for the functioning of biological or engineered systems~\cite{strogatz2004sync}.

Many theoretical studies have been conducted to analyze efficient topologies for achieving stable synchronization within a single network of coupled dynamical systems~\cite{nishikawa2006synchronization,tanaka2008optimal,brede2008synchrony,yanagita2010design,yanagita2012design,skardal2014optimal,montbrio2015macroscopic,burachik2020sparse,palacios2021synchronization}.
The synchronization {\it between} multiple coupled networks of dynamical systems exhibiting collective oscillations has also been analyzed, typically for coupled networks of Kuramoto-type phase oscillators, where all-to-all coupled networks~\cite{okuda1991mutual,montbrio2004synchronization,sheeba2009asymmetry,sheeba2008routes,kawamura2010phase1,kawamura2010phase2,terada2016dynamics,bick2020understanding} or networks with more general topologies~\cite{kori2009collective,montanari2019phase} have been considered.
{The synchronization between networks of chaotic dynamical systems, often in the master-slave configuration, has also been studied~\cite{li2007synchronization,wu2009generalized,boccaletti2014structure}.}

In our previous study~\cite{nakao2018phase}, motivated by experiments on networks of coupled electrochemical oscillators~\cite{wang2000experiments,kiss2002emerging,ocampo2019weak} and studies regarding the collective phase reduction of oscillator populations~\cite{kawamura2008collective,kori2009collective,kawamura2010phase1,kawamura2010phase2,kawamura2011collective}, we generalized the phase reduction theory for limit-cycle oscillators~\cite{kuramoto1984chemical,winfree2001geometry,ermentrout2010mathematical,brown2004phase,nakao2016phase,pietras2019network,kuramoto2019concept,ermentrout2019recent} to networks of coupled dynamical elements exhibiting collective oscillations, where the dynamics of the individual elements and network topologies can be arbitrary.
This method allows us to simplify multidimensional nonlinear equations describing a collectively oscillating network to a one-dimensional equation for the collective phase of the network, provided that external perturbations applied to the network are sufficiently weak. The phase equation takes a simple form characterized by the network's collective frequency and phase sensitivity functions, and it can be used to analyze the synchronization properties of the network in detail.
As a typical application of the phase reduction theory, we analyzed mutual synchronization of two networks connected via a single pair of internetwork connections.

Recently, the optimization of forcing signals for the stable entrainment of a limit-cycle oscillator has been discussed~\cite{moehlis2006optimal,harada2010optimal,dasanayake2011optimal,zlotnik2012optimal,zlotnik2013optimal,tanaka2014optimal,tanaka2015optimal,pikovsky2015maximizing,wilson2015optimal,pyragas2018optimal,kato2020semiclassical}. Similarly, we considered the optimization of mutual coupling between a pair of limit-cycle oscillators, derived an optimal cross-coupling matrix that yielded the maximal linear stability of the in-phase synchronized state under a quadratic constraint on the norm of the coupling matrix~\cite{shirasaka2017optimizing}, and extended the method to spatially extended rhythmic systems~\cite{nakao2014phase,kawamura2017optimizing}. The theory has been further generalized to the optimization of coupling functionals between the oscillators and, for example, an optimal time-delayed coupling for stable synchronization was derived~\cite{watanabe2019optimization}.

In this study, we consider the optimization of an internetwork coupling matrix for a stable in-phase synchronization between a pair of weakly and symmetrically coupled networks of identical properties. 
For the matrix norm, we consider the $L_1$ element-wise norm in addition to the standard {Frobenius} norm.
We demonstrate that two distinct types of optimal internetwork coupling are obtained as a result of the optimization. In the case {with the Frobenius norm}, dense internetwork coupling matrices with relatively small matrix components are obtained. In the $L_1$ case, sparse internetwork coupling matrices are obtained, where only a small number of connections exist.
These results are illustrated via direct numerical simulations of networks of FitzHugh-Nagumo elements.
{
This is the first study to our knowledge that considers optimization of mutual synchronization between collectively oscillating networks 
on the basis of the phase reduction~\cite{nakao2018phase} and sparse optimization~\cite{tibshirani1996regression,hastie2009elements,candes2006robust}.
}

This paper is organized as follows. In Sec. II, we introduce a general model of weakly coupled networks and derive the reduced phase equations. In Sec. III, the formulation of the optimization problems is presented. In Sec. IV, the results {of optimization are illustrated} using two types of networks of FitzHugh-Nagumo elements. Section V summarizes the results and the Appendices provide the numerical data {and further theoretical details.}

\section{Model}

\subsection{Coupled networks}

\begin{figure}[tb]
\centering
\includegraphics[width=0.55\hsize]{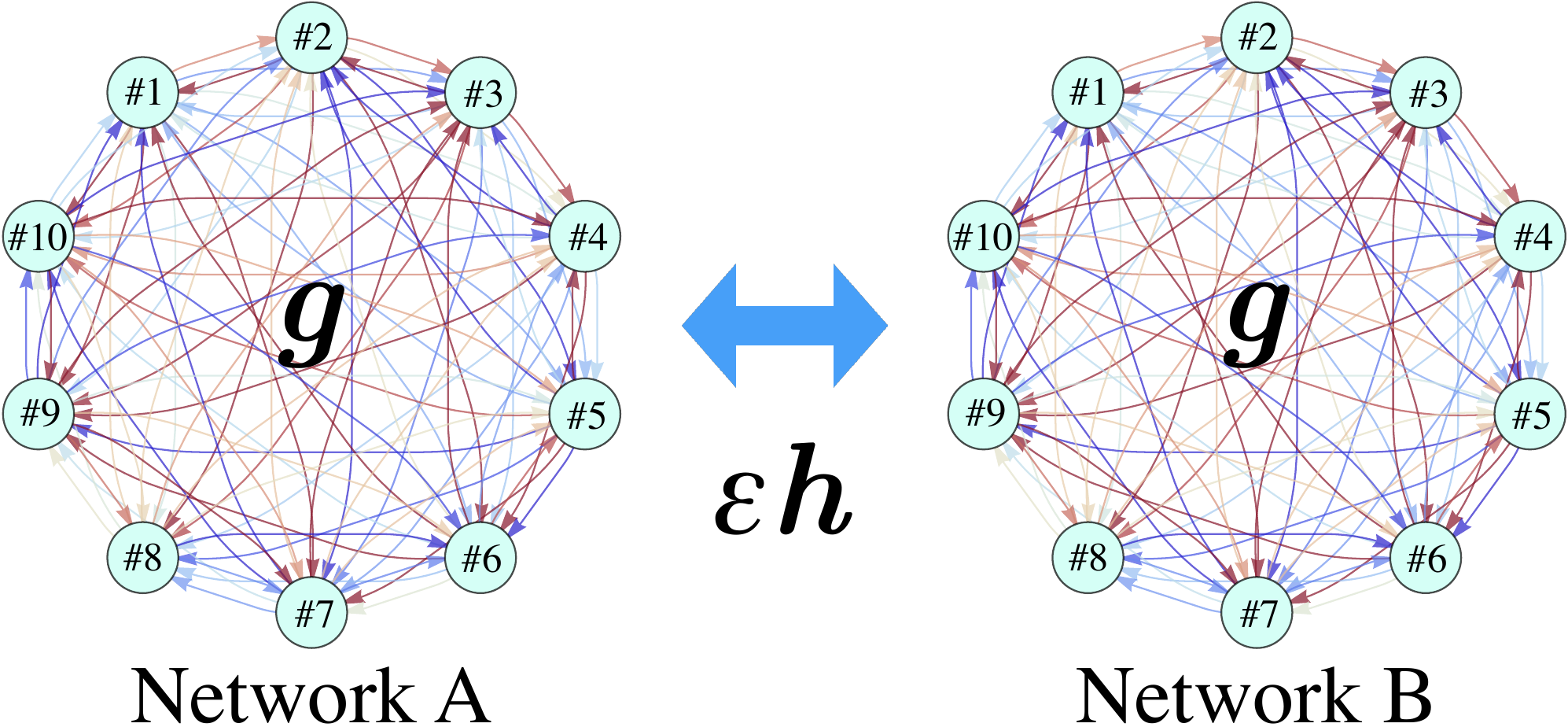}
\caption{A pair of symmetrically coupled networks with identical properties exhibiting stable collective oscillations. It is assumed that the internetwork coupling $\varepsilon {\bm h}$ is much weaker than the intranetwork coupling ${\bm g}$.}
\label{Fig0}
\end{figure}

Figure~\ref{Fig0} shows a schematic illustration of the considered model. We analyze a pair of collectively oscillating networks $A$ and $B$ with identical properties, described by
\begin{align}
	\dot{\bm x}_i^A &= {\bm f}_i({\bm x}_i^A) + \sum_{j=1}^{N} {\bm g}_{ij}({\bm x}_i^A, {\bm x}_j^A) + \varepsilon \sum_{j=1}^{N} {\bm h}_{ij} ( {\bm x}_i^A, {\bm x}_j^B ),
	\cr
	\dot{\bm x}_i^B &= {\bm f}_i({\bm x}_i^B) + \sum_{j=1}^{N} {\bm g}_{ij}({\bm x}_i^B, {\bm x}_j^B) + \varepsilon \sum_{j=1}^{N} {\bm h}_{ij} ( {\bm x}_i^B, {\bm x}_j^A ),
\label{networkeq0}
\end{align}
for $i=1, ..., N$, where $N$ denotes the number of elements in each network, ${\bm x}^{A, B}_i(t) \in {\mathbb R}^{m_i}$ with the dimensionality $m_i \geq 1$ represents the state of element $i$ in the network $A, B$ at time $t$, ${\bm f}_i : {\mathbb R}^{m_i} \to {\mathbb R}^{m_i}$ represents the dynamics of element $i$, ${\bm g}_{ij} : {\mathbb R}^{m_i} \times {\mathbb R}^{m_j} \to {\mathbb R}^{m_i}$ represents intranetwork coupling, i.e., 
the effect that element $i$ is exerted by element $j$
in the same network ($i, j = 1, ..., N$),
${\bm h}_{ij} : {\mathbb R}^{m_i} \times {\mathbb R}^{m_j} \to {\mathbb R}^{m_i}$ represents internetwork coupling, i.e., the effect that element $i$ in one network is exerted by element $j$ in the other network,
and $\varepsilon$ ($0 < \varepsilon \ll 1$) is a small parameter. 
Because the two networks are identical, ${\bm f}_i$, ${\bm g}_{ij}$, and ${\bm h}_{ij}$ are common to both networks.
The state of the network $A, B$ at time $t$ is described by the set $\{ {\bm x}^{A,B}_i(t) \}_{i=1, ..., N} \in {\mathbb R}^{m_1} \times \cdots \times {\mathbb R}^{m_N}$ 
and the dimensionality of each network is $M = \sum_{i=1}^N m_i$.

We assume that, when the two networks are uncoupled ($\varepsilon = 0$),
each network has an exponentially stable limit-cycle solution $\{ \bar{\bm x}_i(t) \}_{i=1, .., N}$ of period $T$ in the $M$-dimensional state space satisfying $\{ \bar{\bm x}_i(t) \}_{i=1, ..., N} = \{ \bar{\bm x}_i(t + T)  \}_{i=1, ..., N}$, corresponding to the stable collective oscillations of the network.
It is noted that the elements can exhibit dynamics that differ from those of the other elements in the network, and that each element may not be oscillatory when isolated; the assumption here is that the network exhibits stable limit-cycle oscillations as a whole (see Figs.~\ref{Fig1} and \ref{Fig2} for examples).

\subsection{Coupled phase equations}

Following the {method of phase reduction for networks}~\cite{nakao2018phase}, we introduce a phase function $\Theta : {\mathbb R}^{m_1} \times \cdots \times {\mathbb R}^{m_N} \to [0, 2\pi)$ for each network, which assigns a phase value $[0, 2\pi)$ to the network state $\{ {\bm x}^{A,B}_i \}_{i=1, ..., N}$ in the basin of the limit cycle such that the phase $\theta^{A,B} = \Theta( \{ {\bm x}^{A,B}_i \}_{i=1, ..., N} )$ of the network always increases with a constant frequency $\omega = 2\pi / T$ when $\varepsilon = 0$.
We represent the network state on the limit cycle as $\{ \bar{\bm x}_i(\theta) \}_{i=1 ..., N}$ as a function of the phase $\theta$ {in the following discussion}. It is noted that by definition, $\Theta(\{\bar{\bm x}_i(\theta)\}_{i=1, ..., N}) = \theta$ holds on the limit cycle.

When $\varepsilon$ is sufficiently small, we can approximate the network state $\{ {\bm x}_i \}_{i=1, ..., N}$ by a nearby state $\{ \bar{\bm x}_i(\theta) \}_{i=1, ..., N}$ on the limit cycle and derive a pair of reduced coupled phase equations from Eq.~(\ref{networkeq0}),
\begin{align}
\dot \theta^{A} &= \omega+\varepsilon \sum_{i=1}^{N} \sum_{j=1}^{N} {\gamma}_{i j}\left(\theta^{A}-\theta^{B}\right) {= \omega + \epsilon \Gamma(\theta^A - \theta^B),}
\cr
\dot \theta^{B} &= \omega+\varepsilon \sum_{i=1}^{N} \sum_{j=1}^{N} {\gamma}_{i j}\left(\theta^{B}-\theta^{A}\right) {= \omega + \epsilon \Gamma(\theta^B - \theta^A),}
\label{networkphase0}
\end{align}
{where }
\begin{align}
{\gamma}_{ij}(\varphi) &= \frac{1}{2 \pi} \int_{0}^{2 \pi} {\bm Q}_{i}(\psi+\varphi) \cdot \boldsymbol{h}_{i j}\left( \bar{\bm x}_{i}(\psi+\varphi), \bar{\bm x}_{j}(\psi) \right) d \psi
\label{phasecouplingfn}
\end{align}
is an {\it individual phase coupling function} describing the effect that {a network receives through its element $i$} from element $j$ in the other network, {and the sum of $\gamma_{ij}$ over all pairs $(i, j)$ of internetwork connections is defined as
\begin{align}
\Gamma(\varphi) = \sum_{i=1}^N \sum_{j=1}^N \gamma_{ij}(\varphi),
\end{align}
which we call the {\it collective phase coupling function}.
}
{In Eq.~(\ref{phasecouplingfn})}, ${\bm Q}_i(\theta)$ ($i=1, ..., N$) is the {\it phase sensitivity function} of element $i$ defined as the gradient $\partial \Theta / \partial {\bm x}_i$ evaluated at state $\{ \bar{\bm x}_i(\theta) \}_{i=1, ..., N}$ with phase $\theta$ on the limit cycle, which characterizes the linear response property of the network phase $\theta$ to weak perturbations applied to element $i$.
It is noted that {${\bm Q}_i$, $\gamma_{ij}$, and thus $\Gamma$} are $2\pi$-periodic functions that are common to both networks.
In Ref.~\cite{nakao2018phase}, it was shown that the phase sensitivity functions $\{ {\bm Q}_i(\theta) \}_{i=1, ..., N}$ can be obtained as $2\pi$-periodic solutions to a set of coupled adjoint-type equations with an appropriate normalization condition 
(see Appendix {A} for a brief description).
{The phase equations obtained above are correct up to $O(\varepsilon)$.}

To analyze the phase synchronization between the networks, we consider the phase difference $\varphi = \theta^A - \theta^B$ between them, which obeys
\begin{align}
\dot{\varphi} 
&= \dot{\theta}^A - \dot{\theta}^B 
{= \epsilon \{ \Gamma(\varphi) - \Gamma(-\varphi) \} = \epsilon \Gamma_a(\varphi)}
\label{phasedifeq}
\end{align}
from Eq.~(\ref{networkphase0}).
Here, we defined the antisymmetric part of the collective phase coupling function $\Gamma(\varphi)$ as
\begin{align}
\Gamma_a(\varphi) = \Gamma(\varphi) - \Gamma(-\varphi).
\end{align}
which describes the evolution of the phase difference $\varphi$ between the networks.

It is clear that the in-phase synchronized state of the two networks, $\varphi = 0$, is always a stationary solution.
The linearization around this solution for a small $| \varphi |$ yields
\begin{align}
\dot{\varphi} 
\approx 
{\epsilon \Gamma_a'(0) \varphi =}
2 \varepsilon \left ( \sum_{i=1}^N \sum_{j=1}^N {\gamma}'_{ij}(0) \right) \varphi.
\end{align}
Hence, the linear stability of the in-phase synchronized state is characterized by
\begin{align}
{\Lambda
= - \Gamma_a'(0) }
= - 2 \sum_{i=1}^N \sum_{j=1}^N {\gamma}_{ij}'(0),
\label{linstab}
\end{align}
which should be positive to achieve the {stable} in-phase synchronization.
The in-phase synchronized state $\varphi=0$ is more stable when {$\Lambda$ is larger}.

{The above results can be straightforwardly generalized to a pair of collectively oscillating nonidentical networks with a small frequency mismatch as described in Appendix B, though we focus on a pair of identical networks for simplicity in what follows.}
 
\subsection{Internetwork coupling matrix}

We assume that the internetwork coupling is expressed as
\begin{align}
{\bm h}_{ij}({\bm x}_i^A, {\bm x}_j^B) = C_{ij} {\bm H}_{{ij}} ({\bm x}_i^A, {\bm x}_j^B),
\quad
(i, j = 1, ..., N),
\label{simplelinear}
\end{align}
where the connection weight $C_{ij}$ is the $i,j$-component of an internetwork coupling matrix $C \in {\mathbb R}^{N \times N}$ and ${\bm H}_{{ij}} : {\mathbb R}^{m_i} \times \mathbb{R}^{m_j} \to \mathbb{R}^{m_i}$ specifies the effect of state vector ${\bm x}_j^B$ of element $j$ in network $B$ on state vector ${\bm x}_i^A$ of element $i$ in network A, and vice versa for ${\bm h}_{ij}({\bm x}_i^B, {\bm x}_j^A)$.
We assume that the function ${\bm H}_{{ij}}$ is specified and fixed, whereas matrix $C$ characterizing the connection weights between the elements can be varied.

The {individual} phase coupling function is expressed as
\begin{align}
{\gamma}_{ij} ( \varphi ) 
&= C_{ij} V_{ij}(\varphi),
\end{align}
for $i, j = 1, ..., N$, where we defined
\begin{align}
V_{ij}(\varphi) 
= \frac{1}{2\pi} \int_{0}^{2\pi} {\bm Q}_i(\psi + \varphi) \cdot {\bm H}_{{ij}} ( \bar{\bm x}_i(\psi + \varphi), \bar{\bm x}_j(\psi) ) d\psi.
\label{V}
\end{align}
Using $V_{ij}(\varphi)$, we can express the linear stability of the synchronized state $\varphi = 0$ as
\begin{align}
\Lambda 
= - 2 \sum_{i=1}^N \sum_{j=1}^N C_{ij} V_{ij}'(0).
\label{linstaV}
\end{align}
Hence, each connection from element $j$ to element $i$ provides a contribution {proportional to} $V_{ij}'(0)$ to the linear stability.

\begin{figure}[tb]
\centering
\includegraphics[width=0.35\hsize]{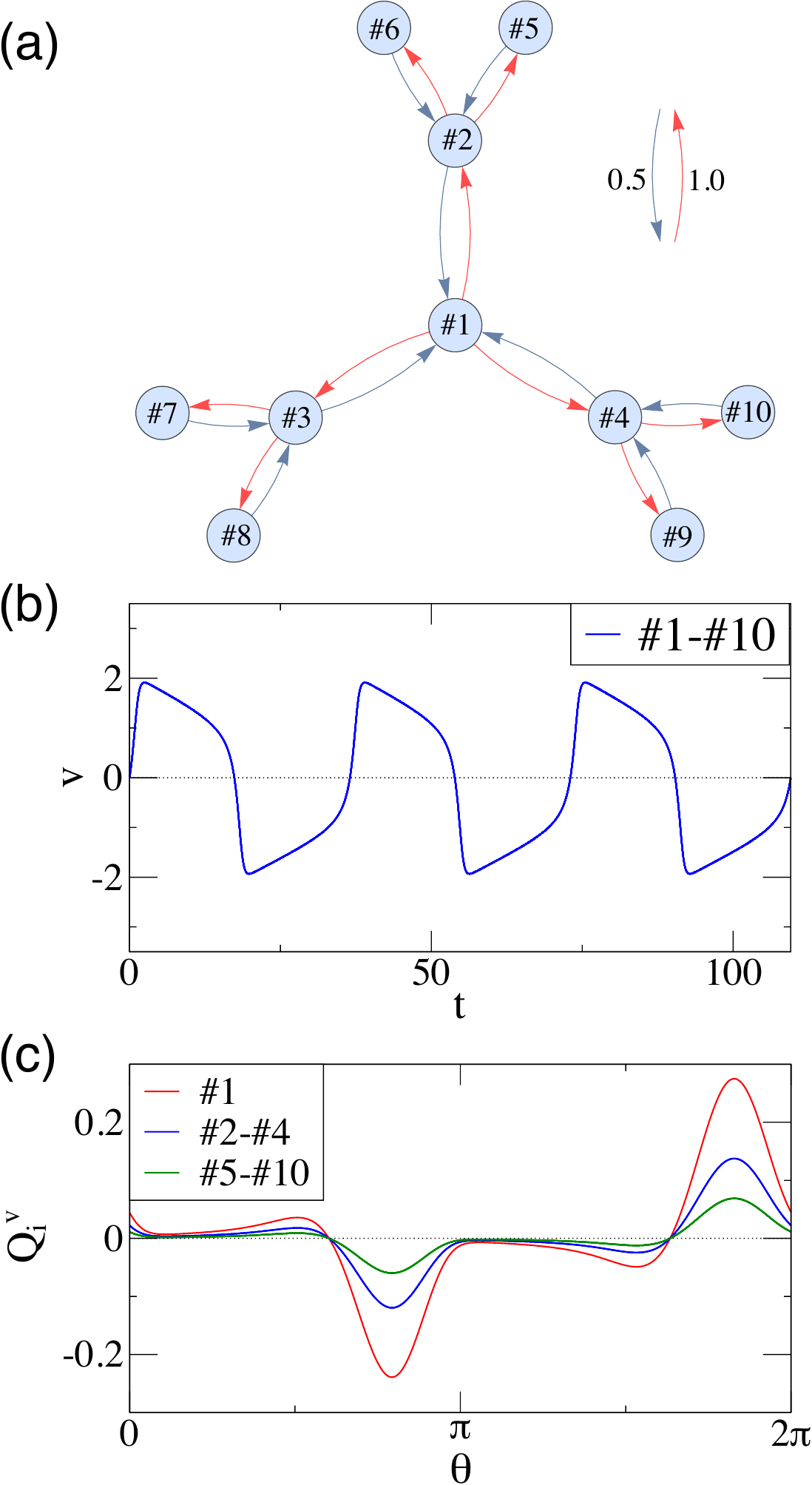}
\caption{(a) Tree network of FitzHugh-Nagumo elements. Circles indicate elements and arrows indicate intranetwork coupling $K$ between elements, where red and blue indicate positive coupling of weights $1.0$ and $0.5$, respectively. (b) Collective oscillations of network. Time evolution of $v$-components $\bar{v}_{1, ..., 10}$ of $\bar{\bm x}_{1, ..., 10}$ are shown for three oscillation periods. All curves overlap with each other because all elements are completely synchronized. (c) $v$-components $Q_{1, ..., 10}^v$ of phase sensitivity functions ${\bm Q}_{1, ..., 10}$ of elements. Depending on the level, three different curves are obtained: $Q_1^v$ for element $1$ at the top level; $Q_{2, 3, 4}^v$ for elements $2, 3, 4$ at the middle level; and $Q_{5, ..., 10}^v$ at the bottom level.}
\label{Fig1}
\end{figure}

\subsection{Optimization of internetwork coupling}

We aim to obtain the most efficient internetwork coupling matrix $C$ for a stable in-phase synchronization between networks.
In evaluating the matrix norm, we consider the standard {Frobenius} norm {(element-wise $L_2$ norm)}, which tends to yield evenly distributed dense internetwork coupling, as well as the element-wise $L_1$ norm, which tends to yield sparse internetwork coupling.
The former case is a network extension of the previous study~\cite{shirasaka2017optimizing}, in which the optimal cross-coupling matrix between the vector components of a pair of weakly coupled limit-cycle oscillators was obtained,
whereas the latter case is newly introduced in this study to consider the sparseness of the internetwork coupling.

Thus, we consider the following optimization problem:
\begin{align}
\max_C \ {\Lambda} \quad \text { s.t. } \quad \|C\|_{{F}, 1} = const.,
\label{optim1}
\end{align}
where $\| ... \|_{F, 1}$ denotes the $L_1$ or {Frobenius} matrix norm.
This maximizes the linear stability ${ \Lambda }$ of the synchronized state $\varphi = 0$ while fixing the norm of $C$ that characterizes the overall intensity of the internetwork coupling.
Alternatively, we may consider the following optimization problem:
\begin{align}
\min_C \ \|C\|_{{F}, 1} \quad \text { s.t. } \quad { \Lambda } = const.
\label{optim2}
\end{align}
Namely, we minimize the overall intensity of the internetwork coupling while fixing the linear stability at a specified value, thereby finding the most efficient coupling between the networks to achieve the prescribed synchronization stability.
Because ${ \Lambda }$ has a simple linear form given by Eq.~(\ref{linstaV}), the optimal solution $C^*$ of the first problem, Eq.~(\ref{optim1}), is proportional to the solution $C^+$ in the second problem, Eq.~(\ref{optim2}).
In the following sections, we consider the second problem and numerically evaluate the optimal $C^+$. Subsequently, we rescale $C^+$ to obtain the optimal $C^*$ for the first problem and use it for numerical illustrations.
\\

\begin{figure}[tb]
\centering
\includegraphics[width=0.65\hsize]{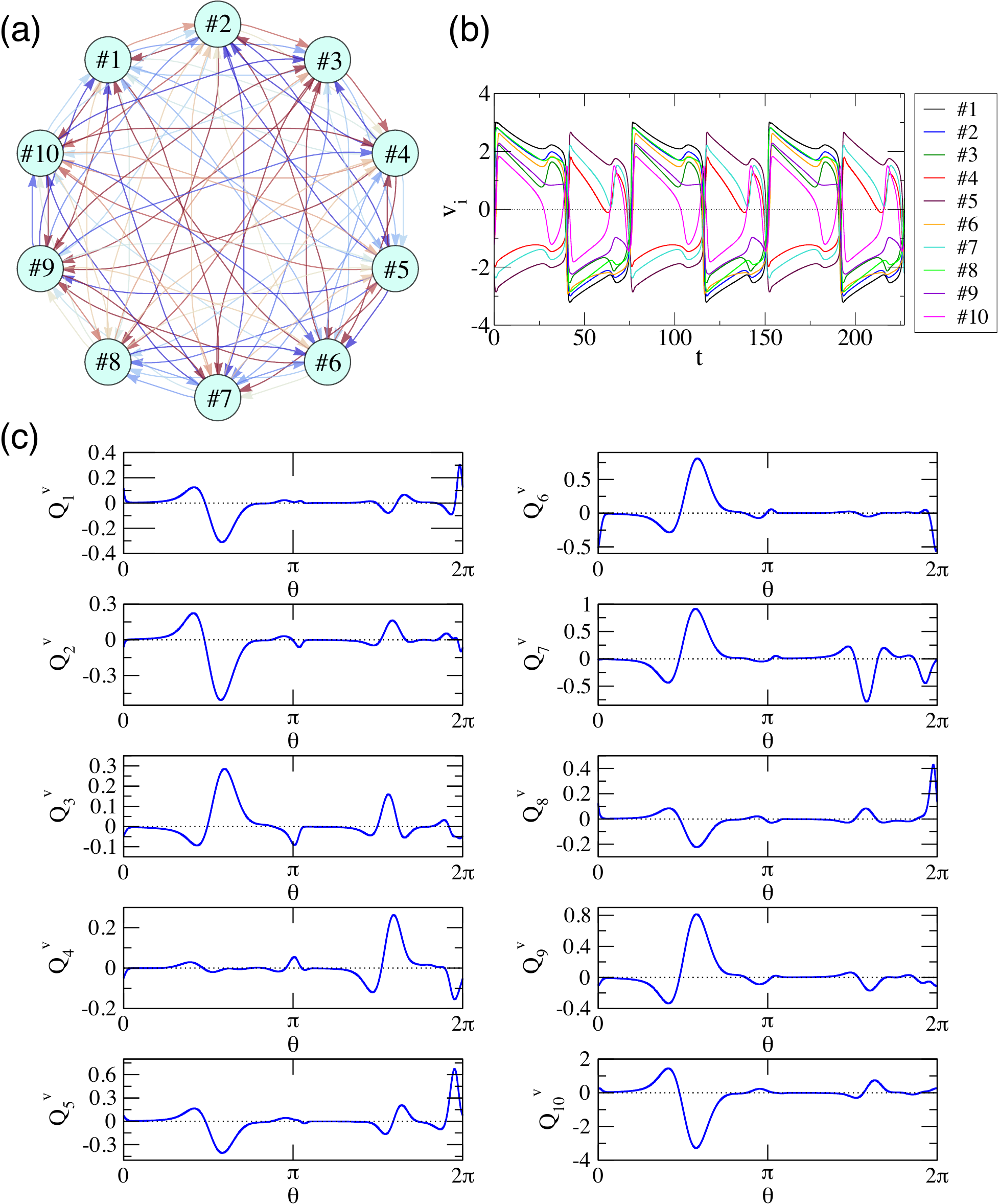}
\caption{(a) Random network of FitzHugh-Nagumo elements. Circles indicate the elements and arrows indicate intranetwork interactions between the elements, where red color indicates positive coupling and blue indicates negative coupling, respectively. (b) Collective oscillations of network. Time evolution of $v$-components $\bar{v}_{1, ..., 10}$ of $\bar{\bm x}_{1, ..., 10}$ are shown in different colors for $3$ periods of oscillations. (c) $v$-components $Q_{1, ..., 10}^v$ of phase sensitivity functions ${\bm Q}_{1, ..., 10}$ of the elements.}
\label{Fig2}
\end{figure}

\section{Optimization}

\subsection{{Frobenius-norm} optimization}

We first consider the Frobenius norm of the coupling matrix $C$,
\begin{align}
\| C \|_{{F}} = \sqrt{ \sum_{i=1}^N \sum_{j=1}^N ( C_{ij} )^2 },
\end{align}
and minimize it under the constraint that the linear stability ${ \Lambda }$ of the in-phase synchronized state, Eq.~(\ref{linstaV}),
is fixed at a given value $q > 0$, i.e.,
\begin{align}
\min_C \ \|C\|_{{F}} \quad \text { s.t. } \quad { \Lambda } = q.
\end{align}

To solve this problem, we introduce a Lagrangian multiplier $\lambda$ and define an objective function as follows:
\begin{align}
S(C, \lambda) &= ( \| C \|_{{F}} )^2 + \lambda ( { \Lambda } - q )
\cr
&= \sum_{i=1}^N \sum_{j=1}^N ( C_{ij} )^2 + \lambda \left( {-} 2 \sum_{i=1}^N \sum_{j=1}^N C_{ij} V_{ij}'(0) - q \right).
\cr
\end{align}
The matrix $C$ and the multiplier $\lambda$ that yield the extremum of $S$ should satisfy
\begin{align}
\frac{\partial S}{\partial C_{ij}}
&= 2 C_{ij} {-} 2 \lambda V_{ij}'(0) = 0
\end{align}
for $i, j = 1, 2, ..., N$, and
\begin{align}
\frac{\partial S}{\partial \lambda} = {-} 2 \sum_{i=1}^N \sum_{j=1}^N C_{ij} V_{ij}'(0) - q = 0.
\end{align}

Hence, the optimal matrix $C$ that minimizes $S$ should be expressed in the form
$C_{ij} = \lambda V_{ij}'(0)$ and the Lagrange multiplier $\lambda$ is given by
\begin{align}
\lambda = - \frac{q}{2 \sum_{i=1}^N \sum_{j=1}^N \left( V_{ij}'(0) \right)^2 } = - \frac{q}{2 ( \| V'(0) \|_{{F}} )^2},
\end{align}
where $V'(0) \in {\mathbb R}^{N \times N}$ is a matrix whose components are specified by $\{ V'_{ij}(0) \}$.
Therefore, the optimal internetwork coupling matrix $C^+_{F}$ with the minimal {Frobenius} norm is given as
\begin{align}
{(C_F^+)}_{ij} = {-}
\frac{q}{2 ( \| V'(0) \|_{{F}}) ^2} V_{ij}'(0),
\end{align} 
where $i, j = 1, ..., N$, and the {Frobenius} norm of $C^+_{F}$ is
\begin{align}
\| C^+_{F} \|_{{F}} = \sqrt{ \sum_{i=1}^N \sum_{j=1}^N \{ (C^+_{F})_{ij} \}^2 } = \frac{ q }{2 \| V'(0) \|_{{F}} }.
\end{align}

Therefore, if we minimize the {Frobenius} norm, the optimal coupling matrix $C^+_{{F}}$ is proportional to the matrix $V'(0)$ defined in Eq.~(\ref{V}).
Because each component of $V'(0)$ is generally non-zero, $C^+_{{F}}$ is a dense matrix in this case; in other words, every element in network $A$ is connected to every element in network $B$, and vice versa.

\subsection{$L_1$-norm optimization}

As described in the previous subsection, we obtained a dense internetwork coupling matrix by minimizing the {Frobenius} norm. 
However, this may not be practical because every pair of elements should be connected.
Therefore, we seek a sparse internetwork coupling matrix.

Borrowing ideas from least absolute shrinkage and selection operator (LASSO) in regression analysis~\cite{tibshirani1996regression,hastie2009elements} and compressed sensing in signal processing~\cite{candes2006robust}, we consider the $L_1$ element-wise norm of $C$,
\begin{align}
\| C \|_1 = \sum_{i=1}^N \sum_{j=1}^N | C_{ij} |,
\end{align}
and minimize $\|C\|_1$ under a specified constraint on the linear stability ${ \Lambda }$.
Because of the diamond-like shapes of the $L_1$-norm level sets~\cite{hastie2009elements}, a sparse matrix $C$ is expected.

However, if we constrain only the linear stability ${ \Lambda }$, we will generally obtain an extremely sparse matrix $C$ as the minimizer, whose components are zero except for a single, large non-zero component.
In other words, $\| C \|_1$ is minimized when a single strong pair of connections between the elements mediates internetwork coupling.
This can be problematic because extremely strong connections between the elements can violate the assumption of weak coupling; furthermore, the internetwork coupling is not resilient because it can easily be destroyed by terminating this pair of connections.
Hence, we introduce a limit $c$ on the absolute value of each matrix component $|C_{ij}|$ representing the connection weight, i.e., $| C_{ij} | \leq c$ for $c > 0$, to avoid a concentrated internetwork coupling on a single pair of connections between the networks.

Therefore, we consider the following minimization problem:
\begin{align}
&\min_C \  \|C\|_{1} 
\cr
&\text { s.t. } \quad
{ \Lambda } = q, \quad | C_{ij} | \leq c \quad (i, j = 1, ..., N),
\end{align}
where ${ \Lambda } $ is expressed by Eq.~(\ref{linstaV}) and $q>0$.
This problem can be cast into the following linear programming form by introducing a new matrix $L = \{ L_{ij} \} \in {\mathbb R}^{N \times N}$:
\begin{align}
&\min \sum_{i=1}^N \sum_{j=1}^N L_{ij}
\cr
&\text { s.t. } 
\cr
&\quad \quad \quad  L_{ij} - C_{ij} \geq 0,\quad L_{ij} + C_{ij} \geq 0,
\cr
&\quad \quad \quad { \Lambda } = q,\ | C_{ij} | \leq c \quad (i, j = 1, ..., N).
\label{l1opt2}
\end{align}
It is noted that $L_{ij} \geq \mbox{max}\{ C_{ij}, -C_{ij} \} = | C_{ij} |$; hence, $ |C_{ij}| = |L_{ij}|$ holds when it is optimized.
The above linear-programming problem can be solved using appropriate numerical solvers~\cite{press2007numerical,Mathematica}.
See Appendix {C} for a brief description of the method used in this study.
{We denote the resulting optimal internetwork coupling matrix by $C_1^+$. }

\subsection{Diagonal coupling}

To evaluate the optimization results, we additionally consider a diagonal internetwork coupling matrix $C^d$ as a typical non-optimized coupling, which is simply proportional to the identity matrix,
\begin{align}
C_{ij}^d = {-} \frac{q}{2} \frac{1}{ \sum_{i=1}^N V'_{ii}(0)} \delta_{i, j},
\end{align}
and yields the same linear stability, where all $N$ pairs of the corresponding elements from the two networks are coupled with each other. The {individual} phase coupling function in this case is expressed as
\begin{align}
{\gamma}_{ij} ( \varphi ) 
= C_{ij}^d V_{ij}(\varphi)
= {-} \frac{q}{2} \frac{1}{\sum_{i=1}^N V'_{ii}(0)} \delta_{i,j} V_{ij}(\varphi) 
\end{align}
for $i, j = 1, ..., N$, and the linear stability of the in-phase state is  $ { \Lambda } = q$.
The {Frobenius} norm of this $C^d$ is expressed as
\begin{align}
\| C^d \|_{{F}} = \sqrt{ \sum_{i=1}^N \sum_{j=1}^N (C^d_{ij})^2 } =  \frac{q}{2} \left|\frac{1}{ \sum_{i=1}^N V'_{ii}(0) } \right| \sqrt{N},
\end{align}
and the $L_1$ norm is expressed as
\begin{align}
\| C^d \|_1 = \sum_{i=1}^N \sum_{j=1}^N \left| C^d_{ij} \right| =  \frac{q}{2} \left| \frac{1}{ \sum_{i=1}^N V'_{ii}(0) } \right| {N}.
\end{align}
We use this $C^d$ as the baseline to evaluate the efficiency of the optimized matrices.
\\

\section{Examples}

\subsection{Networks of FitzHugh-Nagumo elements}

As examples, we use networks of FitzHugh-Nagumo elements for our numerical simulations, as in Ref.~\cite{nakao2018phase}.
Each network consists of $N$ elements
whose state vectors are represented by ${\bm x}_i = (u_i, v_i)^{\sf T} \in {\mathbb R}^2$ for $i=1, ..., N$, where ${\sf T}$ denotes the transpose. The dimensionality of the element is $m_i = 2$ for all $i$ and that of the network is $2N$.
The dynamics of each element is described by
\begin{align}
\boldsymbol{f}_{i}\left({\bm x}_{i}\right)=\left(\delta\left(a+v_{i}-b u_{i}\right), v_{i}-\frac{v_{i}^{3}}{3}-u_{i}+I_{i}\right)^{\sf T},
\end{align}
where $\delta, a, b$, and $I_i$ are parameters. 
We fixed $\delta = 0.08$, $a=0.7$ and $b=0.8$ for the following calculations. 
Depending on $I_i$, each element exhibits oscillatory or {excitable} dynamics.

The intranetwork coupling is assumed to be diffusive and is given by
\begin{align}
\boldsymbol{g}_{i j}\left({\bm x}_{i}, {\bm x}_{j}\right)=K_{i j}\left(0, v_{j}-v_{i}\right)^{\sf T},
\end{align}
where coupling with other elements occurs through the $v$-component, and $K_{ij}$ is the $(i, j)$-component of the $N \times N$ intranetwork coupling matrix $K$ representing the weights of intranetwork diffusive coupling.

Similarly, the internetwork coupling is assumed to be diffusive and is expressed as
\begin{align}
\boldsymbol{h}_{i j}\left({\bm x}_{i}^A, {\bm x}_{j}^B\right)
= C_{ij} {\bm H}_{{ij}}\left({\bm x}_{i}^A, {\bm x}_{j}^B\right)
= C_{ij} \left(0, v_{j}-v_{i}\right)^{\sf T}.
\end{align}
In this case, the {individual} phase coupling function is given by
\begin{align}
{\gamma}_{ij} ( \varphi ) = C_{ij} V_{ij}(\varphi),
\end{align}
with
\begin{align}
V_{ij}(\varphi) 
&= \frac{1}{2\pi} \int_{0}^{2\pi} Q_i^{v}(\psi+\varphi) \{ \bar{v}_j(\psi) - \bar{v}_i(\psi+\varphi) \} d\psi,
\label{Vconv}
\end{align}
where $Q_i^{v}$ is the $v$-component of ${\bm Q}_i$ and $\bar{v}_j$ is the $v$-component of $\bar{\bm x}_j$.

\begin{figure}[tb]
\centering
\includegraphics[width=0.4\hsize]{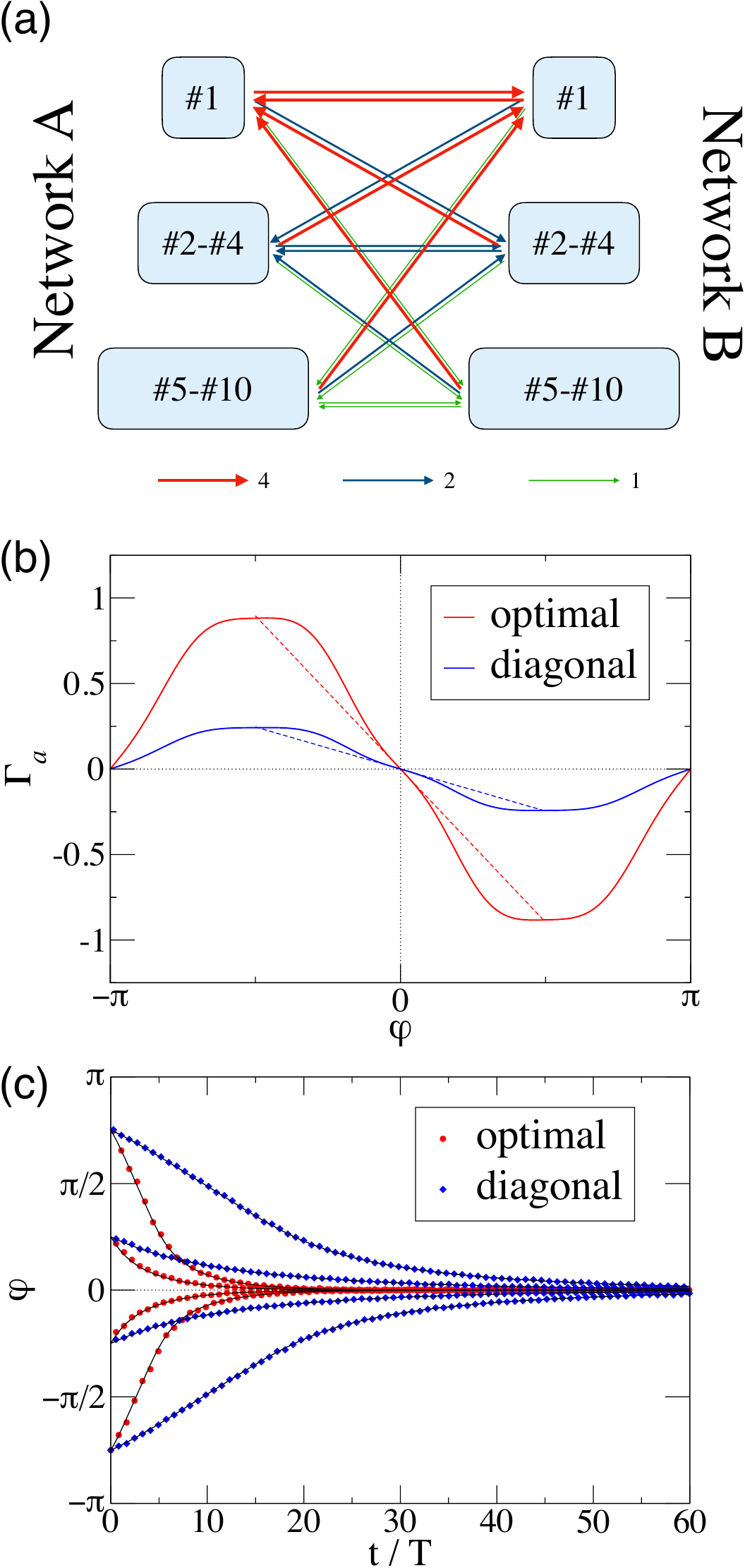}
\caption{{Frobenius-norm} optimization for tree networks. (a) Optimal internetwork coupling $C^+_{F}$. Colors of arrows represent relative weights of coupling (thick red: $4$, middle blue: $2$, thin green: $1$). Intranetwork coupling $K$ is not shown. (b) {Antisymmetric part of the collective} phase coupling functions for optimal and diagonal internetwork coupling with rescaled connection weights. Dotted lines represent slopes at the origin, characterizing (negative) linear stability of in-phase synchronized state. (c) Evolution of phase difference $\varphi$ for optimal and diagonal internetwork coupling with rescaled connection weights. Symbols show results of direct numerical simulations, and solid curves show results of reduced phase equations. }
\label{fig3}
\end{figure}

\subsection{Network topologies and collective oscillations}

We consider two types of networks consisting of $N=10$ FitzHugh-Nagumo elements.\\

(i) Tree network (Fig.~\ref{Fig1}(a); see Appendix {D} for the intranetwork coupling matrix $K$). This network comprises three levels (top: element $\#1$; middle: elements $\#2-\#4$; bottom: elements $\#5-\#10$), and the coupling from higher to lower-level elements is twice as strong as the coupling in the opposite directions. All elements have identical parameter values $I_i = 0.8$ ($i=1, ..., 10$) and are self-oscillatory. This network has a stable limit-cycle solution of period $T \approx 36.5$ and frequency $\omega \approx 0.172$ where all elements are completely synchronized, i.e., $\bar{\bm x}_1 = ... = \bar{\bm x}_{10}$, as shown in Fig.~\ref{Fig1}(b). 

Figure~\ref{Fig1}(c) shows the $v$-components of $\bar{\bm x}_{1, ..., 10}$ and ${\bm Q}_{1, ..., 10}$.
Reflecting the three levels of the network and asymmetric coupling among them, the phase sensitivity functions of the elements are classified into three cases and the elements at the same level share the same phase sensitivity function, namely, ${\bm Q}_1$ for element $\#1$, ${\bm Q}_2 = {\bm Q_3} = {\bm Q}_4$ for elements $\#2-\#4$, and ${\bm Q}_5 = ... = {\bm Q}_{10}$ for elements $\#5-\#10$. In addition, reflecting the simple topology, all phase sensitivity functions have the same shape and only their amplitudes differ based on the levels, satisfying ${\bm Q}_1 = 2 {\bm Q}_{2, 3, 4} = 4 {\bm Q}_{5-10}$. Hence, element $\#1$ at the top level can be regarded as a dynamical hub of the network in that perturbations applied to this element imposes the most significant effect on the network's collective oscillations.
{See Appendix E for the matrix $V'(0)$ obtained numerically for this tree network.}

(ii) Random network (Fig.~\ref{Fig2}(a); see Appendix {D} for the intranetwork coupling matrix $K$). This network is the same as that used in Ref.~\cite{nakao2018phase}. Elements $\#1-\#8$ are {excitable} with $I_{1, ..., 8} = 0.2$  and elements $\#9$ and $\#10$ are oscillatory with $I_{9,10} = 0.8$. Each component of the intranetwork connection weight $K$ is randomly drawn from $[-0.6, 0.6]$.

This network has a stable limit cycle with period $T \approx 75.7$ and frequency $\omega \approx 0.0830$, where each individual element undergoes the same periodic dynamics in unison with all other elements. The dynamics of the elements are all different owing to the random network coupling and heterogeneity of parameter $I_i$.

Figure~\ref{Fig2}(c) shows the $v$-components of the phase sensitivity functions ${\bm Q}_{1, ..., 10}$ of the network.
The results shown are the same as those obtained in Ref.~\cite{nakao2018phase}.
It is observed that element $\#10$ has a phase sensitivity function $Q_{10}^v$ with a considerably larger amplitude than the other elements and can be regarded as a dynamical hub of the network.
{The matrix $V'(0)$ obtained numerically for this random network is given in Appendix E.}

\begin{figure}[tb]
\centering
\includegraphics[width=0.4\hsize]{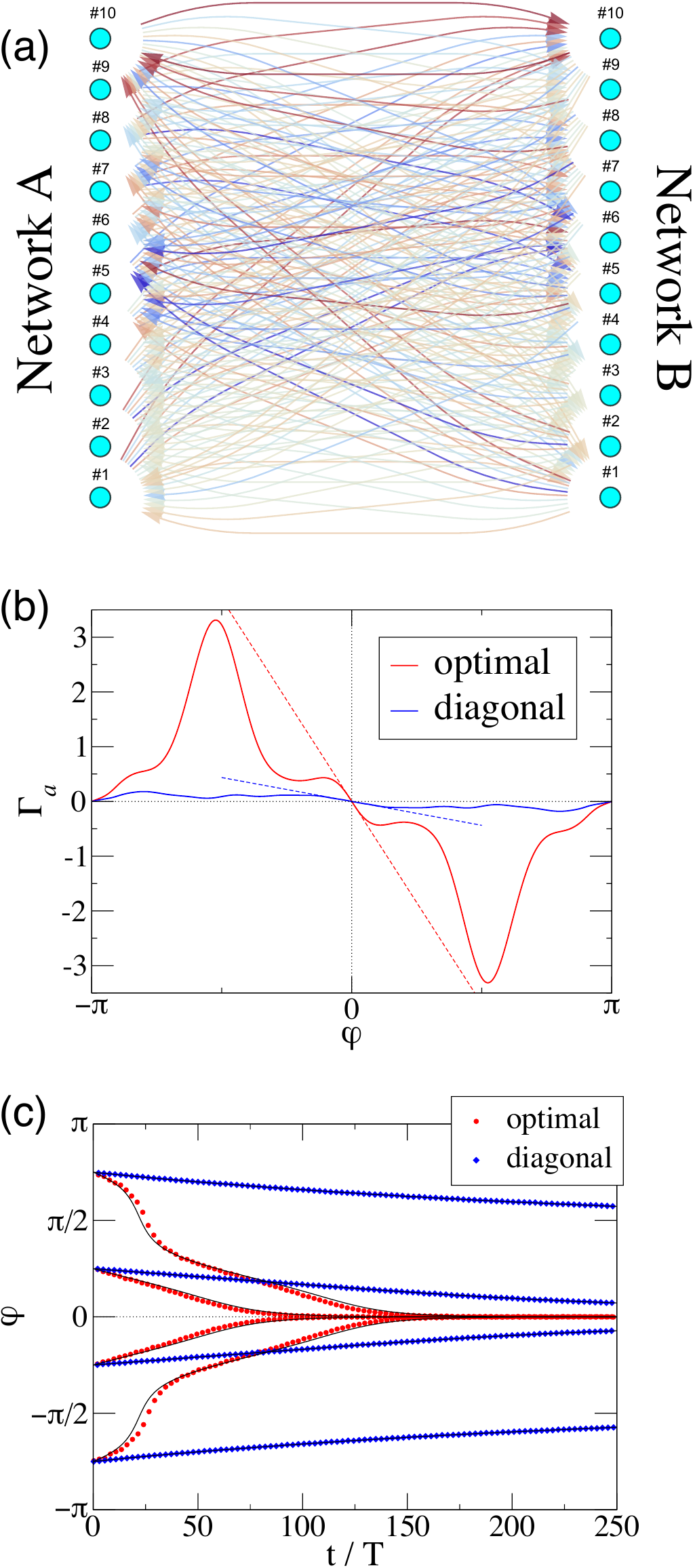}
\caption{{Frobenius-norm} optimization for random networks. (a) Optimal internetwork coupling $C^+_{F}$. Red and blue arrows indicate positive and negative interactions, respectively. Intranetwork coupling $K$ in each network is not shown. (b) {Antisymmetric part of the collective} phase coupling functions for optimal and diagonal internetwork coupling with rescaled connection weights. Dotted lines represent slopes at the origin. (c) Evolution of phase difference $\varphi$ for optimal and diagonal internetwork coupling with rescaled connection weights. Symbols show results of direct numerical simulations, and solid curves show results of reduced phase equations. }
\label{fig4}
\end{figure}

\subsection{{Frobenius-norm} optimization}

We first present the results of {Frobenius-norm} optimization of internetwork coupling for the tree and random networks. We solved the second optimization problem, Eq.~(\ref{optim2}), namely, we fixed the linear stability ${ \Lambda }$ at $q$, where we assumed $q=1$, and calculated the optimized coupling matrix $C^+_{F}$ and the corresponding diagonal matrix $C^d$.
Next, we normalized $C^+_{F}$ as ${C}^*_{F} = C^+_{F} / \| C^+_{F} \|_{{F}}$, thereby solving the first optimization problem, Eq.~(\ref{optim1}) with $\| C^*_{F} \|_{{F}} = 1$. Similarly, we normalized $C^d$ as $C^D = C^d / \| C^d \|_{{F}}$.
We used $C^*_{F}$ and $C^D$ to calculate the {collective} phase coupling function and performed numerical simulations so that the optimization results can be demonstrated as an improvement in the linear stability. 
\\

(i) Tree networks. Figure~\ref{fig3}(a) shows the optimal internetwork coupling $C^+_{{F}}$ between networks $A$ and $B$ schematically, where the elements are classified into three levels (top: $\#1$; middle: $\#2-\#4$; bottom: $\#5-\#10$), and the arrows indicate the directions and weights of the internetwork coupling between the elements at the respective levels. In this case, all matrix components of $C^+_{F}$ were positive. See Appendix {F} for the actual $C^+_{F}$. 

Because all elements in each network are completely synchronized and exhibit the same dynamics, the elements at the same level received an internetwork coupling of equal weight from all the elements in the other network.
As indicated by the arrows with different colors, hub element $\#1$ at the top level received $4X$ stronger coupling, whereas elements $\#2-\#4$ in the middle received $2X$ stronger coupling compared with elements $\#5-\#10$ at the bottom level.
The {Frobenius} norm of the optimized coupling matrix $C^+_{{F}}$ was $\| C^+_{{F}} \|_{{F}} \approx 1.75$, whereas that of the diagonal coupling matrix $C^d$ yielding the same ${ \Lambda }$ was $\| C^d \|_{{F}} \approx 6.37$.
Thus, the norm of the optimized $C^+_{{F}}$ was considerably smaller than that of the diagonal $C^d$.

Next, we normalized $C^+_{{F}}$ by the norm $\| C^+_{{F}} \|_{{F}}$ to define $C^*_{{F}} = C^+_{{F}} / \| C^+_{{F}} \|_{{F}}$ and used this $C^*_{{F}}$ to calculate the {collective} phase coupling function and performed numerical simulations. Similarly, we normalized the diagonal $C^d$ and defined ${C}^D = C^d / \| C^d \|_{{F}}$.

Figure~\ref{fig3}(b) shows the antisymmetric part of the {collective} phase coupling functions $\Gamma_a(\varphi) = \Gamma(\varphi) - \Gamma(-\varphi)$ calculated for the normalized optimal matrix $C^*_{{F}}$ and normalized diagonal matrix $C^D$. As shown, the optimal $C^*_{{F}}$ yielded a larger linear stability, given by the (negative) slope of $\Gamma_a(\varphi)$ at $\varphi = 0$, compared with the diagonal ${C}^D$ of the same norm.
The linear stability ${ \Lambda }$ was $0.570$ and $0.157$ for $C^*_{{F}}$ and $C^D$, respectively.

Figure~\ref{fig3}(c) shows the evolution of the phase differences $\varphi = \theta^A - \theta^B$ between the networks for cases with the optimal $C^*_{{F}}$ and diagonal ${C}^D$, where the results of direct numerical simulations (DNS) of Eq.~(\ref{networkeq0}) are compared with those obtained using the phase equation~(\ref{phasedifeq}). The small parameter $\varepsilon$ was fixed at $0.01$.
The optimized coupling provided a faster convergence to the in-phase state compared with the diagonal coupling, and the results of the phase model agreed well with those of the DNS (slight deviations were due to the approximation error in the phase reduction). 
\\

(ii) Random networks. Figure~\ref{fig4}(a) shows the optimal internetwork coupling between the networks. See Appendix {F} for the actual $C^+_{{F}}$. In this case, the matrix components of $C^+_{{F}}$ could take both positive and negative values. As explained previously, the internetwork coupling was dense and all pairs of elements were coupled between the networks. We observed that hub node $\#10$ in each network obtained relatively strong connections from the elements in the other network.
The {Frobenius} norm of the optimal $C^+_{{F}}$ was $\| C^+_{{F}} \|_{{F}} \approx 0.424$, whereas that of the diagonal $C^d$ yielding the same linear stability was $\| C^d \|_{{F}} \approx 3.59$.

Figure~\ref{fig4}(b) shows the antisymmetric part of the {collective} phase coupling function $\Gamma_a(\varphi)$ for the normalized optimal and diagonal matrices, $C^*_{{F}}$ and $C^D$, respectively. The shapes of $\Gamma_a$ differed considerably between the two cases, and the slope of $\Gamma_a$ at the origin was much larger for $C^*_{{F}}$ than for $C^D$.
The linear stability ${ \Lambda }$ was $2.36$ for $C^*_{{F}}$ and $0.278$ for $C^D$, respectively.
Figure~\ref{fig4}(c) shows the evolution of the phase difference $\varphi$ between the networks obtained via DNS and the phase model for $\varepsilon = 0.0003$. The optimal $C^*_{{F}}$ yielded a much faster convergence to the synchronized state $\varphi=0$ than the diagonal $C^D$. 

Thus, efficient internetwork coupling matrices were obtained via {Frobenius-norm} optimization for the two types of networks. In both tree and random networks, the hub-like elements in those networks tend to gain stronger internetwork coupling than the other elements; however, concentrating the coupling on a small number of connections results in a higher {Frobenius-norm} cost. The balance between these effects yields the optimal internetwork coupling, which densely connects all pairs of elements between the networks.

\begin{figure}[tb]
\centering
\includegraphics[width=0.75\hsize]{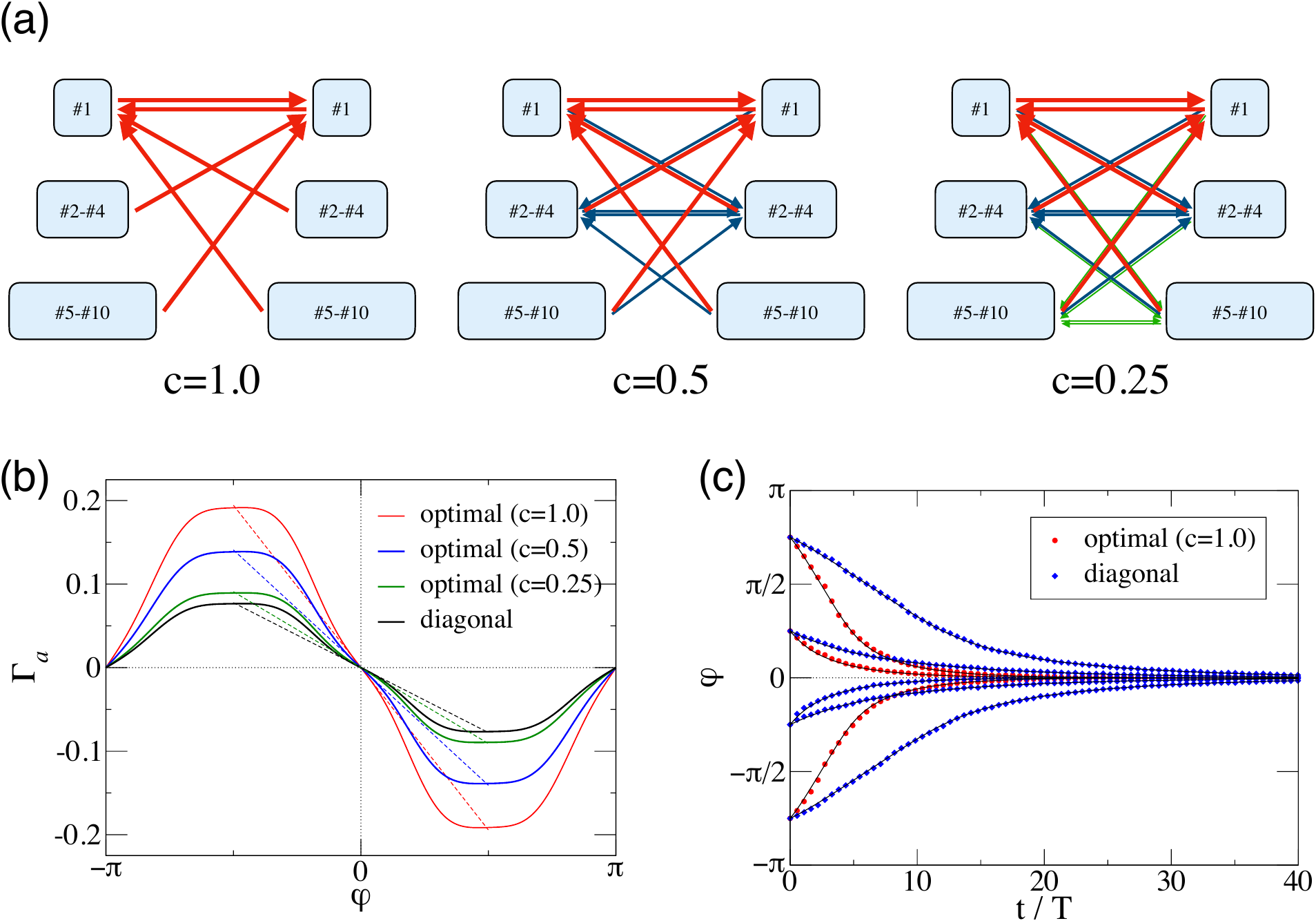}
\caption{$L_1$-norm optimization for tree networks. (a) Optimal internetwork coupling {$C^+_{1}$} for $c=1.0$, $0.5$, and $0.25$. Red, blue, and green arrows indicate coupling to the top, middle, and bottom levels, respectively. (b) {Antisymmetric part of the collective} phase coupling functions for optimal ($c=1.5$, $1.0$, and $0.5$) and diagonal internetwork coupling with rescaled connection weights. Dotted lines represent slopes at the origin. (c) Evolution of phase difference $\varphi$ for optimal ($c=1.5$) and diagonal internetwork coupling with rescaled connection weights. Symbols show results of direct numerical simulations, and solid curves show results of reduced phase equations. }
\label{fig5}
\end{figure}

\subsection{$L_1$-norm optimization}

We next present the results of $L_1$-norm optimization of the internetwork coupling matrices for tree and random networks. We fixed the linear stability ${ \Lambda }$ as $q=1$ as in the previous case and additionally introduced the limit $c$ on $|C_{ij}|$.
\\

(i) Tree networks. Figure~\ref{fig5}(a) shows a schematic illustration of the optimized internetwork coupling $C^+_{1}$ 
for $c=1.0$, $0.5$, and $0.25$ (see Appendix {G} for the matrix components). When $c=1.0$, a sufficiently large pairwise coupling was allowed and only element $\#1$ at the top level received the internetwork coupling. Because all elements in each network exhibit the same dynamics, all connections had the same weight. When $c=0.5$, the coupling to element $\#1$ was limited, and the elements $\#2-\#4$ at the middle level received additional coupling to achieve the required linear stability. When $c$ was further reduced to $0.25$, elements $\#5-\#10$ at the bottom level also received coupling. The norm of $C^+_{1}$ was approximately $8.06$, $11.1$, and $17.2$ for $c=1.0$, $0.5$, and $0.25$, whereas that of $C^D$ was $20.2$.

It should be noted that the result above is not unique. Because the dynamics of all elements in the network are identical and the elements at the same level share the same {individual} phase sensitivity function, the matrix components of $V'(0)$ in the same rows are identical to each other {(Appendix E)}. 
Therefore, we can increase or decrease the connection weights to the elements of the same level such that their sum remains the same to obtain different $C^+_{1}$ while maintaining the same linear stability and $L_1$ norm. The optimal solution shown here is the symmetric case where all the connection weights are the same.

Figure~\ref{fig5}(b) shows the antisymmetric part of the {collective} phase coupling functions $\Gamma_a$ calculated using the three normalized optimal coupling matrices $C^*_{1}$ for $c=1.0$, $0.5$, and $0.25$ as well as using the normalized diagonal matrix $C^D$. As shown, the optimal $C^*_{1}$ with $c=1.0$ yielded the most significant improvement, whereas $C^*_{1}$ with $c=0.25$ yielded only a small improvement compared with the diagonal $C^D$. The linear stability ${ \Lambda }$ was approximately $0.124$, $0.0901$, and $0.0581$ for $C^*_{1}$ with $c=1.0$, $0.5$, and $0.25$, respectively, and $0.0496$ for $C^D$.

Figure~\ref{fig5}(c) shows the evolution of the phase differences obtained via DNS and the phase equation using the optimal $C^*_{1}$ for $c=1.0$ and the diagonal $C^D$ with $\epsilon=0.05$. We observed a faster convergence to the in-phase synchronized state $\varphi=0$ in the optimized case. 
\\

(ii) Random networks. Figure~\ref{fig6}(a) shows the optimized internetwork coupling $C^+_{1}$ for $c=1.5$, $1.0$, and $0.5$ (see Appendix {G} for the matrix components). For $c=1.5$, all elements of $C^+_{1}$ were zero except for a single component $(C^+_{1})_{10,10}$, which corresponded to the sparsest case where only elements $\#10$ of both networks were coupled. When $c=1.0$, an additional pair of connections from elements $\#1$ to $\#6$ emerged because of the limitation, and when $c=0.5$, another additional pair of connections from $\#8$ to $\#6$ appeared. The number of connections increased as $c$ was further decreased, and the optimization problem became unsolvable when $c$ was too small.
In contrast to case (i) for the tree networks, these solutions were unique because all matrix components of $V'(0)$ differed from each other {(Appendix E)}.
The norm of $C^+_{1}$ was approximately $1.31$, $1.35$, and $1.48$ for $c=1.5$, $1.0$, and $0.5$, respectively, whereas that of $C^D$ was $11.3$.
Thus, sparse internetwork coupling matrices were obtained via the $L_1$-norm optimization.

Figure~\ref{fig6}(b) shows the antisymmetric part of the {collective} phase coupling functions $\Gamma_a$ for the three normalized optimal coupling matrices $C^*_{1}$ with $c=1.5$, $1.0$, and $0.5$ and for the normalized diagonal matrix $C^D$. The optimal cases yielded much higher linear stability, where ${ \Lambda } = 0.763, 0.740$, and $0.676$ for $C^*_{1}$ with $c=1.0$, $0.5$, and $0.25$, respectively, and ${ \Lambda=0.0881 }$ for $C^D$.
It is noteworthy that ${ \Lambda}$ showed only a small decrease with a decrease in $c$ (i.e., an increase in the number of connections) in this range. Therefore, by selecting the appropriate $c$, we can obtain a sparse internetwork coupling matrix but still with several connections that gives a nearly maximal linear stability while affording resilience to connection failure between the networks.

Figure~\ref{fig6}(c) shows the evolution of the phase differences obtained via DNS and the phase equation using the optimal $C^*_{1}$ for $c=1.5$ and the diagonal $C^D$ for $\epsilon=0.002$. A much faster convergence to the in-phase synchronized state was observed in the optimized case.
\\
\begin{figure}[tb]
\centering
\includegraphics[width=0.75\hsize]{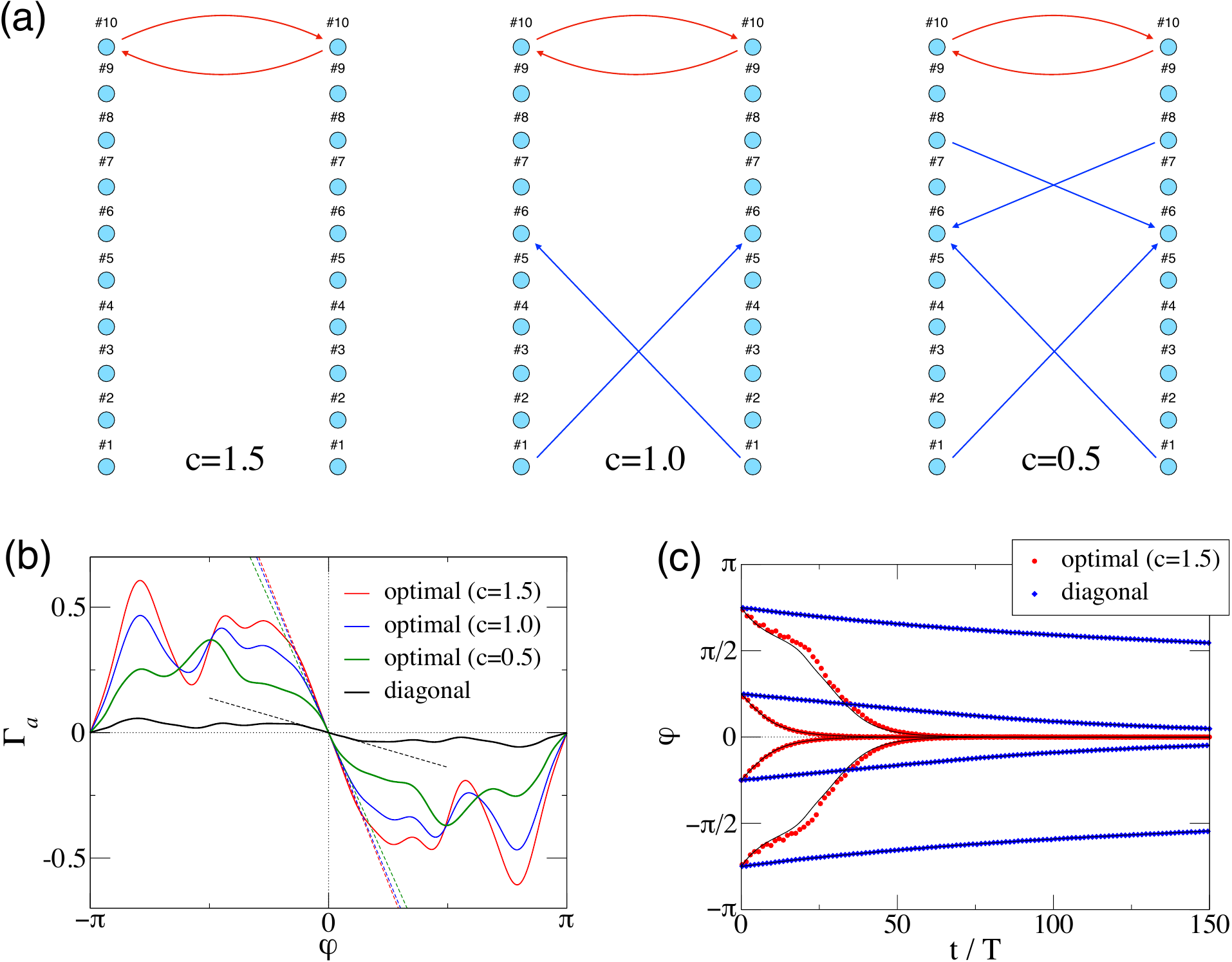}
\caption{$L_1$-norm optimization for random networks. (a) Optimal internetwork coupling  {$C^+_{1}$} for $c=1.5$, $1.0$, and $0.5$. Red and blue arrows indicate positive and negative coupling, respectively. (b) {Antisymmetric part of the collective} phase coupling functions for optimal ($c=1.5$, $1.0$, and $0.5$) and diagonal internetwork coupling matrices with rescaled connection weights. Dotted lines represent slopes at the origin. (c) Evolution of phase difference $\varphi$ for optimal ($c=1.5$) and diagonal internetwork coupling matrices with rescaled connection weights. Symbols show results of direct numerical simulations and solid curves show results of reduced phase equations. }
\label{fig6}
\end{figure}

\section{Concluding remarks}

We considered two types of optimization problems for the internetwork coupling between a pair of coupled networks exhibiting stable collective oscillations and obtained optimal internetwork coupling matrices that achieved maximal linear stability in the in-phase synchronized state.
For the {Frobenius} norm, dense internetwork coupling matrices were obtained, whereas for the $L_1$ norm, sparse internetwork coupling matrices were obtained. The sparseness of the internetwork coupling can be controlled by introducing additional constraints on the connection weights, thereby yielding sparse yet resilient internetwork connections.

{In Appendix H, we summarize the results of the two types of optimization, where the norms of the coupling matrices for a given linear stability and the linear stability attained with the normalized coupling matrices are shown in the tables for the tree and random networks. It is interesting that the $L_1$-norm optimization with additional weight constraints can yield sparse coupling matrices that are also reasonably good in terms of the Frobenius norm.}

Theoretically, the most important quantity is the matrix $V'(0)$ whose $i, j$-component characterizes the effect of element $j$ on element $i$ obtained by a convolution of the phase sensitivity ${\bm Q}_i^v$ with the waveform $v_j - v_i$ of the diffusive coupling, as given by Eq.~(\ref{Vconv}). 
It is notable that the results of $L_1$-norm optimization can be understood intuitively. Namely, with the decrease in the limit $c$ on the connection weights, the optimization selects the connections between the elements from the most important pairs one by one, where the importance is determined by the absolute value of the matrix component of $|V'(0)|$.
This is because the linear stability ${ \Lambda}$ is determined by a simple linear sum in the form of Eq.~(\ref{linstaV}) and the greedy algorithm provides the exact solution.
The results of {Frobenius-norm} optimization can also be understood intuitively, namely, the relation $C_{ij} \propto V'_{ij}(0)$ indicates that
the vector $\{C_{ij}\}$ is parallel to $\{V'_{ij}(0)\}$ and 
each connection obtains a connection weight that is proportional to its importance on synchronization, yielding a minimal $\| C \|_{{F}}$.

{In this study, we considered a pair of symmetrically coupled identical networks for simplicity. In this case, $\phi=0$ and $\phi=\pi$ are always the fixed points of Eq.~(\ref{phasedifeq}). In general, there can also be other stable fixed points and the networks can synchronize with non-zero phase difference (e.g., in  Fig. 4(d) of Ref.~\cite{nakao2018phase}, four stable fixed points exist). 
Also, even if the coupled networks with the diagonal internetwork coupling have $\phi = 0$ and $\pi$ as the only fixed points, the networks may possess other fixed points when coupled with the optimized coupling matrices. 
As our aim is to improve the linear stability of the in-phase synchronized state $\phi=0$, we did not consider such cases in the present study.
Though stable fixed points other than $\phi=0$ did not occur for the networks and coupling matrices used in our examples, if the global stability of the fixed point $\phi = 0$ is required, we need to include additional constraints explicitly in the optimization to avoid the existence of unnecessary fixed points.
Though not considered in the present study, we recently introduced such a constraint on the global convergence property in the optimization of periodic inputs for the entrainment of a limit-cycle oscillator~\cite{kato2021optimization}, and we would be able to perform similar optimization of coupled networks such that global convergence to $\varphi=0$ is ensured.}

The present results are based on the phase reduction theory and it can be extended to more general cases, e.g., to several collectively oscillating networks with different properties, provided that the networks exhibit stable collective oscillations and the internetwork coupling is sufficiently weak. 
{See Appendix B for the derivation of the phase equations for the case with nonidentical networks with a small frequency mismatch.}
For non-weak internetwork coupling, the phase reduction becomes inaccurate and the present results may not apply. Recently, a phase-amplitude reduction theory that considers deviations of the system state from the limit cycle due to non-weak perturbations has been developed~\cite{wilson2016isostable,shirasaka2017phase,ermentrout2019recent,kuramoto2019concept}, and it has been used to devise non-weak periodic inputs to achieve fast entrainment~\cite{nakao2021pobo,takata2021fast}. {Such a phase-amplitude framework would also be applicable to coupled networks.}

Finally, from a practical viewpoint, the dense network obtained via {Frobenius-norm} optimization may be difficult to implement because every pair of elements should be connected, whereas sparse networks with several connections obtained via $L_1$-norm optimization would be easier to implement.
This result may provide useful insights into the design of networked dynamical systems with efficient synchronization properties.

\begin{acknowledgments}
H.N. acknowledges JSPS KAKENHI JP17H03279, JP18H03287, JP18K03471, JPJSBP120202201, and JST CREST JP-MJCR1913 for financial support.\\
\end{acknowledgments}

\begin{center}
{\bf Data availability}
\end{center}

{The data that support the findings of this study are available within the article.}\\


\appendix

\section{Adjoint equation for the phase sensitivity function}

The phase sensitivity function ${\bm Q}_i$ of element $i$ is given by the gradient of the phase function $\Theta$ with respect to ${\bm x}_i$
evaluated at the network state $\{ \bar{\bm x}_i(\theta) \}_{i=1, ..., 10}$ on the limit cycle, ${\bm Q}_i(\theta) = ( \partial \Theta / \partial {\bm x}_i )_{ \{ \bar{\bm x}_i(\theta) \}_{i=1, ..., N}}$.
In Ref.~\cite{nakao2018phase}, it is shown that the function ${\bm Q}_i(\theta)$ is a $2\pi$-periodic solution to the following coupled adjoint-type equations:
\begin{align}
\omega \frac{d}{d\theta} {\bm Q}_i(\theta) = - J_i(\theta)^{\sf T} {\bm Q}_i(\theta) - \sum_{j=1}^N M_{ij}(\theta)^{\sf T} {\bm Q}_i(\theta) - \sum_{j=1}^N N_{ij}(\theta)^{\sf T} {\bm Q}_j(\theta)
\quad
(i=1, ..., N),
\end{align}
where 
\begin{align}
J_i(\theta) = \left. \frac{\partial {\bm f}_i({\bm x}_i)}{\partial {\bm x}_i} \right|_{\{\bar{\bm x}_i(\theta)\}_{i=1, ..., N}},
\quad
M_{ij}(\theta) = \left. \frac{\partial {\bm g}_{ij}({\bm x}_i, {\bm x}_j)}{\partial {\bm x}_i} \right|_{\{\bar{\bm x}_i(\theta)\}_{i=1, ..., N}}
\end{align}
are $m_i \times m_i$ real matrices, and
\begin{align}
N_{ij}(\theta) = \left. \frac{\partial {\bm g}_{ij}({\bm x}_i, {\bm x}_j)}{\partial {\bm x}_i} \right|_{\{\bar{\bm x}_j(\theta)\}_{i=1, ..., N}}
\end{align}
is a $m_i \times m_j$ real matrix representing the Jacobian of ${\bm f}_i$ and ${\bm g}_i$ evaluated at $\{\bar{\bm x}_i(\theta)\}_{i=1, ..., N}$ at phase $\theta$ on $\chi$. A normalization condition
\begin{align}
\sum_{j=1}^N {\bm Q}_i(\theta) \cdot \frac{d \bar{\bm x}_i(\theta)}{d \theta} = 1
\end{align}
should also be satisfied to be consistent with the definition of the asymptotic phase.
The adjoint equations above can be solved numerically via backward integration with occasional normalization~\cite{ermentrout2010mathematical}.

\section{Coupled nonidentical networks}

In the main text, we focus on a pair of symmetrically coupled identical networks for simplicity, but the theory can be straightforwardly generalized to coupled nonidentical networks with a small frequency mismatch. Here, we briefly present a derivation of the phase equations for such a case.
We consider two weakly coupled nonidentical networks, each of which exhibiting stable limit-cycle oscillations, described by
\begin{align}
	\dot{\bm x}_i^A &= {\bm f}_i^A({\bm x}^A_i) + \sum_{j=1}^{N_A} {\bm g}^A_{ij}({\bm x}_i^A, {\bm x}_j^A) + \varepsilon \sum_{j=1}^{N_B} {\bm h}^A_{ij} ( {\bm x}_i^A, {\bm x}_j^B )
	\quad (i=1, ..., N^A),
	\cr
	\dot{\bm x}_i^B &= {\bm f}_i^B({\bm x}^B_i) + \sum_{j=1}^{N_B} {\bm g}^B_{ij}({\bm x}_i^B, {\bm x}_j^B) + \varepsilon \sum_{j=1}^{N_A} {\bm h}^B_{ij} ( {\bm x}_i^B, {\bm x}_j^A )
	\quad (i=1, ..., N^B),
\label{networkeqg}
\end{align}
where $N^{A,B}$ denotes the number of elements in network $A,B$, ${\bm x}^{A,B}_i(t) \in {\mathbb R}^{m^{A,B}_i}$ with the dimensionality $m^{A,B}_i \geq 1$ represents the state of element $i$ in network $A, B$ at time $t$, ${\bm f}^{A,B}_i : {\mathbb R}^{m^{A,B}_i} \to {\mathbb R}^{m^{A,B}_i}$ represents dynamics of element $i$ in network $A, B$, and ${\bm g}^{A,B}_{ij} : {\mathbb R}^{m^{A,B}_i} \times {\mathbb R}^{m^{A,B}_j} \to {\mathbb R}^{m^{A,B}_i}$ represents intranetwork coupling within network $A, B$.
Moreover, 
${\bm h}^{A}_{ij} : {\mathbb R}^{m^{A}_i} \times {\mathbb R}^{m^{B}_j} \to {\mathbb R}^{m^{A}_i}$
and
${\bm h}^{B}_{ij} : {\mathbb R}^{m^{B}_i} \times {\mathbb R}^{m^{A}_j} \to {\mathbb R}^{m^{B}_i}$
represent internetwork connections between the networks $A$ and $B$,
where 
${\bm h}^{A}_{ij}$ represents the effect that element $j$ in network $B$ exerts on element $i$ in network $A$, 
and
${\bm h}^{B}_{ij}$ represents the effect that element $j$ in network $A$ exerts on element $i$ in network $B$, 
respectively,
and
$\varepsilon$ ($0 < \varepsilon \ll 1$) is a small parameter. 
The state of the network $A, B$ at time $t$ is described by the set $\{ {\bm x}^{A,B}_i(t) \}_{i=1, ..., N^{A,B}} \in {\mathbb R}^{m^{A,B}_1} \times \cdots \times {\mathbb R}^{m^{A,B}_{N^{A,B}}}$ 
and the dimensionality of each network is $M^{A,B} = \sum_{i=1}^{N^{A,B}} m^{A,B}_i$.

We assume that when the two networks are uncoupled ($\varepsilon = 0$), each network has an exponentially stable limit-cycle solution $\{ \bar{\bm x}^{A,B}_i(t) \}_{i=1, .., N^{A,B}}$ of period $T^{A,B}$ in the $M^{A,B}$-dimensional state space satisfying $\{ \bar{\bm x}^{A,B}_i(t) \}_{i=1, ..., {N^{A,B}}} = \{ \bar{\bm x}^{A,B}_i(t + T^{A,B})  \}_{i=1, ..., N^{A,B}}$, corresponding to the stable collective oscillations of the network.
We introduce a phase function $\Theta^{A,B} : {\mathbb R}^{m^{A,B}_1} \times \cdots \times {\mathbb R}^{m^{A,B}_{N^{A,B}}} \to [0, 2\pi)$ for each network, which assigns a phase value $[0, 2\pi)$ to the network state $\{ {\bm x}^{A,B}_i \}_{i=1, ..., N^{A,B}}$ in the basin of the limit cycle such that the phase $\theta^{A,B} = \Theta^{A,B}( \{ {\bm x}^{A,B}_i \}_{i=1, ..., {N^{A,B}}} )$ of the network $A, B$ always increases with a constant frequency $\omega^{A,B} = 2\pi / T^{A,B}$ when $\varepsilon = 0$.
We represent the state of each network on the limit cycle as $\{ \bar{\bm x}^{A,B}_i(\theta) \}_{i=1 ..., N^{A,B}}$ as a function of the phase $\theta$
and introduce the phase sensitivity functions of each element in each network as
\begin{align}
{\bm Q}^{A, B}_i(\theta) = \left. \frac{\partial \Theta^{A,B}}{\partial {\bm x}^{A,B}_i} \right|_{\{ {\bm x}^{A,B}_i = \bar{\bm x}^{A,B}_j(\theta) \}_{j=1, ..., N^{A,B}}\}}
\quad
(i=1, ..., N^{A,B}),
\end{align}
which can be calculated numerically by solving the adjoint equation for the network $A, B$.

When $\varepsilon$ is sufficiently small, we can approximate the network state $\{ {\bm x}^{A,B}_i \}_{i=1, ..., N^{A,B}}$ by a nearby state $\{ \bar{\bm x}^{A,B}_i(\theta^{A,B}) \}_{i=1, ..., N^{A,B}}$ on the limit cycle and derive a pair of reduced coupled phase equations from Eq.~(\ref{networkeqg}) as
\begin{align}
\dot \theta^{A} &= \omega^A+\varepsilon \sum_{i=1}^{N^A} \sum_{j=1}^{N^B} {\bm Q}^A_{i}(\theta^{A}) \cdot \boldsymbol{h}^A_{i j}\left( \bar{\bm x}^A_{i}(\theta^{A}), \bar{\bm x}^B_{j}(\theta^{B}) \right) ,
\cr
\dot \theta^{B} &= \omega^B+\varepsilon \sum_{i=1}^{N^B} \sum_{j=1}^{N^A} {\bm Q}^B_{i}(\theta^{B}) \cdot \boldsymbol{h}^B_{i j}\left( \bar{\bm x}^B_{i}(\theta^{B}), \bar{\bm x}^A_{j}(\theta^{A}) \right).
\label{networkphaseg}
\end{align}
We assume that the frequency mismatch between $\omega^A$ and $\omega^B$ of the two networks is small and of order $O(\varepsilon)$, and express them as
$\omega^{A,B} = \omega + \varepsilon \Delta^{A,B}$
where $\omega$ is common to both networks and the $\Delta^{A,B}$ is of order $O(1)$.
Introducing slow phase variables $\phi^{A,B} = \theta^{A,B} - \omega t$, we obtain
\begin{align}
\dot \phi^{A} &= \varepsilon \left[ \Delta^A+ \sum_{i=1}^{N^A} \sum_{j=1}^{N^B} {\bm Q}^A_{i}(\omega t + \phi^{A}) \cdot \boldsymbol{h}^A_{i j}\left( \bar{\bm x}^A_{i}(\omega t + \phi^{A}), \bar{\bm x}^B_{j}(\omega t + \phi^{B}) \right) \right],
\cr
\dot \phi^{B} &= \varepsilon \left[ \Delta^B+ \sum_{i=1}^{N^B} \sum_{j=1}^{N^A} {\bm Q}^B_{i}(\omega t + \phi^{B}) \cdot \boldsymbol{h}^B_{i j}\left( \bar{\bm x}^B_{i}(\omega t + \phi^{B}), \bar{\bm x}^A_{j}(\omega t + \phi^{A}) \right) \right].
\end{align}
As the right-hand sides of these equations are of order $O(\varepsilon)$, we employ the averaging approximation, i.e., we average the interaction terms over one period of fast oscillation of $\psi = \omega t$ from $0$ to $2\pi$ and define the {\it individual phase coupling functions}
\begin{align}
{\gamma}^A_{ij}(\phi^A - \phi^B) &= \frac{1}{2 \pi} \int_{0}^{2 \pi} {\bm Q}^A_{i}(\psi+\phi^A) \cdot \boldsymbol{h}^A_{i j}\left( \bar{\bm x}^A_{i}(\psi+\phi^A), \bar{\bm x}^B_{j}(\psi + \phi^B) \right) d \psi
\end{align}
for $i=1, ..., N^A$ and $j=1 ..., N^B$, and
\begin{align}
{\gamma}^B_{ij}(\phi^B - \phi^A) &= \frac{1}{2 \pi} \int_{0}^{2 \pi} {\bm Q}^B_{i}(\psi+\phi^B) \cdot \boldsymbol{h}^B_{i j}\left( \bar{\bm x}^B_{i}(\psi+\phi^B), \bar{\bm x}^A_{j}(\psi + \phi^A) \right) d \psi
\end{align}
for $i=1, ..., N^B$ and $j=1 ..., N^A$. It is noted that all quantities involved in the above equations are $2\pi$-periodic.
The approximate equations for the slow phase variables are 
\begin{align}
\dot{\phi}^A = \varepsilon \left[ \Delta^A + \sum_{i=1}^{N^A} \sum_{j=1}^{N^B} \gamma^A_{ij}(\phi^A - \phi^B) \right],
\cr
\dot{\phi}^B = \varepsilon \left[ \Delta^B + \sum_{i=1}^{N^B} \sum_{j=1}^{N^A} \gamma^B_{ij}(\phi^B - \phi^A) \right],
\end{align}
and, in the original phase variables, we obtain
\begin{align}
\dot{\theta}^A = \omega^A + \varepsilon \sum_{i=1}^{N^A} \sum_{j=1}^{N^B} \gamma^A_{ij}(\theta^A - \theta^B) = \omega^A + \varepsilon \Gamma^A(\theta^A - \theta^B),
\cr
\dot{\theta}^B = \omega^B + \varepsilon \sum_{i=1}^{N^B} \sum_{j=1}^{N^A} \gamma^B_{ij}(\theta^B - \theta^A) = \omega^B + \varepsilon \Gamma^B(\theta^B - \theta^A),
\end{align}
where we defined the sums of individual phase coupling functions over all pairs of the internetwork connections as
\begin{align}
\Gamma^A(\varphi) = \sum_{i=1}^{N^A} \sum_{j=1}^{N^B} \gamma^A_{ij}(\varphi),
\quad
\Gamma^B(\varphi) = \sum_{i=1}^{N^B} \sum_{j=1}^{N^A} \gamma^B_{ij}(\varphi),
\end{align}
which we call the {\it collective phase coupling functions} for the networks $A$ and $B$.

The phase difference $\varphi = \theta^A - \theta^B$ between the two networks is given by
\begin{align}
\dot{\varphi} = ( \omega^A - \omega^B ) + \varepsilon \Gamma_a( \varphi ) = \epsilon \left[ ( \Delta^A - \Delta^B ) + \Gamma_a(\varphi) \right],
\label{phasedifg}
\end{align}
where
\begin{align}
\Gamma_a(\varphi) = \Gamma^A(\varphi) - \Gamma^B(-\varphi)
\end{align}
is the difference between the sums of the collective phase coupling functions (not necessarily antisymmetric in the present case as the networks are nonidentical).
The two networks mutually synchronize when Eq.~(\ref{phasedifg}) has a stable fixed point at some $\varphi^*$, and the linear stability exponent of this synchronized state with phase difference $\varphi^*$ is given by
\begin{align}
\Gamma_a'(\varphi^*) 
= \sum_{i=1}^{N^A} \sum_{j=1}^{N^B} \gamma^A_{ij}(\varphi^*) - \sum_{i=1}^{N^B} \sum_{j=1}^{N^A} \gamma^B_{ij}(-\varphi^*).
\end{align}
It is noted that $\varphi = 0, \pi$ may not be the fixed points as the networks are not identical, and there can also be more than two stable fixed points.

Thus, by defining the linear stability as $\Lambda = - \Gamma_a'(\varphi^*)$ as in Eq.~(\ref{linstab}), we can generally formulate optimization of coupling matrices between the two nonidentical networks $A$ and $B$ with a small frequency mismatch by extending the argument given in the main text.
The above results go back to those in the main text when the two networks are identical and the fixed point at $\varphi^* = 0$ is considered.

\section{Linear programming}

We briefly describe the method to solve the $L_1$-norm optimization problem given by Eq.~(\ref{l1opt2}) numerically.
Representing the $N \times N$ matrices $L$, $C$, and $V'(0)$ as $N^2$-dimensional vectors ${\bm L}$, ${\bm C}$ and ${\bm V}$, we can write the problem as
\begin{align}
&\min {\bm e}^{\sf T} {\bm L}  \quad \text { s.t. } \quad {\bm L} \geq {\bm C}, \quad {\bm L} \geq -{\bm C}, \quad 2 {\bm V}^{\sf T} {\bm C} = - q, \quad {\bm C} \geq - c {\bm e}, \quad {\bm C} \leq c {\bm e},
\end{align}
where $q > 0$ and $c > 0$ are given constants, ${\bm e}$ is an $N^2$-dimensional column vector whose components are all $1$, and the inequalities apply individually to each vector component. Introducing $2N^2$-dimensional column vectors ${\bm E}$ and ${\bm X}$, a $4N^2 \times 2N^2$ matrix $M$, a $4N^2$-dimensional column vector ${\bm B}$, and a $1 \times 2N^2$ matrix (row vector) $N$ as
\begin{align}
{\bm E} = ( \underbrace{1, ..., 1}_{N^2}, \ \underbrace{0, ..., 0}_{N^2} )^{\sf T},
\quad
{\bm X} = ( {\bm L}^{\sf T}, \ {\bm C}^{\sf T} )^{\sf T},
\quad
M = \begin{pmatrix} I & -I \\ I & I \\ 0 & I \\ 0 & -I \end{pmatrix},
\cr
{\bm B} = ( \underbrace{0, ..., 0}_{N^2}, \ - c {\bm e}^{\sf T}, \ - c {\bm e}^{\sf T} ),
\quad
N = ( \underbrace{0, ..., 0}_{N^2}, \ { 2 {\bm V}^{\sf T} } ),
\end{align}
we can further rewrite the problem above as
\begin{align}
\min {\bm E}^{\sf T} {\bm X} \quad \mbox{s.t.} \quad M {\bm X} \geq {\bm B}, \quad N {\bm X} = - q,
\end{align}
which has the standard form of linear programming and can be solved by using appropriate numerical solvers.
In this study, we used the "LinearProgramming" command in Mathematica~\cite{Mathematica} to solve the problem above.

\section{Intranetwork coupling}

The following intranetwork coupling matrices $K$ are used as examples. The random network is the same as that used in Ref.~\cite{nakao2018phase}.

\subsection{Tree network}

\begin{align}
K=
\begin{pmatrix}
0 & 0.5 & 0.5 & 0.5 & 0 & 0 & 0 & 0 & 0 & 0 \\
1.0 & 0 & 0 & 0 & 0.5 & 0.5 & 0 & 0 & 0 & 0 \\
1.0 & 0 & 0 & 0 & 0 & 0 & 0.5 & 0.5 & 0 & 0 \\
1.0 & 0 & 0 & 0 & 0 & 0 & 0 & 0 & 0.5 & 0.5 \\
0 & 1.0 & 0 & 0 & 0 & 0 & 0 & 0 & 0 & 0 \\
0 & 1.0 & 0 & 0 & 0 & 0 & 0 & 0 & 0 & 0 \\
0 & 0 & 1.0 & 0 & 0 & 0 & 0 & 0 & 0 & 0 \\
0 & 0 & 1.0 & 0 & 0 & 0 & 0 & 0 & 0 & 0 \\
0 & 0 & 0 & 1.0 & 0 & 0 & 0 & 0 & 0 & 0 \\
0 & 0 & 0 & 1.0 & 0 & 0 & 0 & 0 & 0 & 0 
\end{pmatrix}
\end{align}

\subsection{Random network}

\begin{align}
K
\approx
\begin{pmatrix}
0.000 & 0.409 & -0.176 & -0.064 & -0.218 & 0.464 & -0.581 & 0.101 & -0.409 & -0.140 \\
0.229 & 0.000 & 0.480 & -0.404 & -0.409 & 0.040 & 0.125 & 0.099 & -0.276 & -0.131 \\
-0.248 & 0.291 & 0.000 & -0.509 & -0.114 & 0.429 & 0.530 & 0.195 & 0.416 & -0.597 \\
-0.045 & 0.039 & 0.345 & 0.000 & 0.579 & -0.232 & 0.121 & 0.130 & -0.345 & 0.463 \\
-0.234 & -0.418 & -0.195 & -0.135 & 0.000 & 0.304 & 0.124 & 0.038 & -0.049 & 0.183 \\
-0.207 & 0.536 & -0.158 & 0.533 & -0.591 & 0.000 & -0.273 & -0.571 & 0.110 & -0.354 \\
0.453 & -0.529 & -0.287 & -0.237 & 0.470 & -0.002 & 0.000 & -0.256 & 0.438 & 0.211 \\
-0.050 & 0.552 & 0.330 & -0.148 & -0.326 & -0.175 & -0.240 & 0.000 & 0.263 & 0.079 \\
0.389 & -0.131 & 0.383 & 0.413 & -0.383 & 0.532 & -0.090 & 0.025 & 0.000 & 0.496 \\
0.459 & 0.314 & -0.121 & 0.226 & 0.314 & -0.114 & -0.450 & -0.018 & -0.333 & 0.000
\end{pmatrix}
\end{align}

\section{Matrix $V'(0)$}

The following matrices were numerically obtained as $V'(0)$ in Eq.~(\ref{V}) and used for the optimization of the internetwork coupling matrices.

\subsection{Tree network}

\begin{align}
V'(0)
\approx
\begin{pmatrix}
-0.0620 & -0.0620 & -0.0620 & -0.0620 & -0.0620 & -0.0620 & -0.0620 & -0.0620 & -0.0620 & -0.0620 \\
-0.0310 & -0.0310 & -0.0310 & -0.0310 & -0.0310 & -0.0310 & -0.0310 & -0.0310 & -0.0310 & -0.0310 \\
-0.0310 & -0.0310 & -0.0310 & -0.0310 & -0.0310 & -0.0310 & -0.0310 & -0.0310 & -0.0310 & -0.0310 \\
-0.0310 & -0.0310 & -0.0310 & -0.0310 & -0.0310 & -0.0310 & -0.0310 & -0.0310 & -0.0310 & -0.0310 \\
-0.0155 & -0.0155 & -0.0155 & -0.0155 & -0.0155 & -0.0155 & -0.0155 & -0.0155 & -0.0155 & -0.0155 \\ 
-0.0155 & -0.0155 & -0.0155 & -0.0155 & -0.0155 & -0.0155 & -0.0155 & -0.0155 & -0.0155 & -0.0155 \\ 
-0.0155 & -0.0155 & -0.0155 & -0.0155 & -0.0155 & -0.0155 & -0.0155 & -0.0155 & -0.0155 & -0.0155 \\ 
-0.0155 & -0.0155 & -0.0155 & -0.0155 & -0.0155 & -0.0155 & -0.0155 & -0.0155 & -0.0155 & -0.0155 \\ 
-0.0155 & -0.0155 & -0.0155 & -0.0155 & -0.0155 & -0.0155 & -0.0155 & -0.0155 & -0.0155 & -0.0155 \\ 
-0.0155 & -0.0155 & -0.0155 & -0.0155 & -0.0155 & -0.0155 & -0.0155 & -0.0155 & -0.0155 & -0.0155 
\end{pmatrix}
\end{align}

The diagonal internetwork coupling matrix yielding the linear stability ${\Lambda = q = 1}$ is $C^d \approx 2.015 I$, where $I$ is a $10 \times 10$ identity matrix and $\| C^d \|_{{F}} \approx 6.372$.  {$\| C^d \|_1 \approx 20.15$, respectively.}
\\

\subsection{Random network}

\begin{align}
V'(0)
\approx
\begin{pmatrix}
-0.084 &-0.063 &-0.013 &0.022 &0.112 &-0.022 &0.053 &-0.082 &-0.042 &-0.068 \\
0.016 &0.002 &-0.029 &0.003 &-0.022 &-0.024 &0.012 &0.009 &-0.013 &-0.010 \\
0.028 &0.020 &-0.004 &-0.002 &-0.028 &0.000 &-0.013 &0.028 &0.021 &0.051 \\
0.044 &0.041 &0.035 &-0.046 &-0.057 &0.021 &-0.052 &0.037 &0.004 &-0.048 \\
-0.138 &-0.116 &-0.070 &0.161 &0.184 &-0.056 &0.243 &-0.124 &-0.038 &0.021 \\
0.337 &0.291 &0.132 &-0.075 &-0.264 &0.123 &-0.125 &0.296 &0.162 &0.143 \\
0.085 &0.087 &0.087 &-0.141 &-0.104 &0.065 &-0.172 &0.082 &0.030 &-0.064 \\
-0.133 &-0.110 &-0.057 &0.101 &0.182 &-0.056 &0.155 &-0.123 &-0.042 &-0.003 \\
0.086 &0.078 &0.044 &-0.028 &-0.069 &0.045 &-0.054 &0.083 &0.061 &0.077 \\
-0.235 &-0.209 &-0.115 &0.023 &0.179 &-0.120 &0.129 &-0.245 &-0.224 &-0.381 
\end{pmatrix}
\end{align}

The diagonal internetwork coupling matrix yielding the linear stability ${\Lambda = q = 1}$ is
$C^d \approx 1.135 I$,
where $I$ is a $10 \times 10$ identity matrix and $\| C^d \|_{{F}} \approx 3.589$. {$\| C^d \|_1 \approx 11.35$}.
\\

\section{{Frobenius-norm}-optimized coupling matrices}

The following matrices were obtained as the {Frobenius-norm}-optimal internetwork coupling matrices $C^+_{F}$ for the tree and random networks. The linear stability was fixed at ${\Lambda = q = 1}$.

\subsection{Tree network}

\begin{align}
C^+_{F}
\approx
\left(
\begin{array}{cccccccccc}
0.3793 & 0.3793 & 0.3793 & 0.3793 & 0.3793 & 0.3793 & 0.3793 & 0.3793 & 0.3793 & 0.3793 \\
0.1897 & 0.1897 & 0.1897 & 0.1897 & 0.1897 & 0.1897 & 0.1897 & 0.1897 & 0.1897 & 0.1897 \\
0.1897 & 0.1897 & 0.1897 & 0.1897 & 0.1897 & 0.1897 & 0.1897 & 0.1897 & 0.1897 & 0.1897 \\
0.1897 & 0.1897 & 0.1897 & 0.1897 & 0.1897 & 0.1897 & 0.1897 & 0.1897 & 0.1897 & 0.1897 \\
0.0948 & 0.0948 & 0.0948 & 0.0948 & 0.0948 & 0.0948 & 0.0948 & 0.0948 & 0.0948 & 0.0948 \\
0.0948 & 0.0948 & 0.0948 & 0.0948 & 0.0948 & 0.0948 & 0.0948 & 0.0948 & 0.0948 & 0.0948 \\
0.0948 & 0.0948 & 0.0948 & 0.0948 & 0.0948 & 0.0948 & 0.0948 & 0.0948 & 0.0948 & 0.0948 \\
0.0948 & 0.0948 & 0.0948 & 0.0948 & 0.0948 & 0.0948 & 0.0948 & 0.0948 & 0.0948 & 0.0948 \\
0.0948 & 0.0948 & 0.0948 & 0.0948 & 0.0948 & 0.0948 & 0.0948 & 0.0948 & 0.0948 & 0.0948 \\
0.0948 & 0.0948 & 0.0948 & 0.0948 & 0.0948 & 0.0948 & 0.0948 & 0.0948 & 0.0948 & 0.0948
\end{array}
\right)
\end{align}
{The Frobenius and $L_1$ norms of this coupling matrix are $\| C^+_{F} \|_{{F}} \approx 1.749$ and $\| C^+_{F} \|_{1}  \approx 15.17$, respectively.}
\\

\subsection{Random network}

\begin{align}
C^+_{F} \approx
\left(
\begin{array}{cccccccccc}
0.030 & 0.023 & 0.005 & -0.008 & -0.040 & 0.008 & -0.019 & 0.029 & 0.015 & 0.024 \\
-0.006 & -0.001 & 0.010 & -0.001 & 0.008 & 0.009 & -0.004 & -0.003 & 0.005 & 0.004 \\
-0.010 & -0.007 & 0.002 & 0.001 & 0.010 & -0.000 & 0.005 & -0.010 & -0.008 & -0.018 \\
-0.016 & -0.015 & -0.013 & 0.017 & 0.021 & -0.007 & 0.019 & -0.013 & -0.001 & 0.017 \\
0.049 & 0.042 & 0.025 & -0.058 & -0.066 & 0.020 & -0.087 & 0.045 & 0.014 & -0.008 \\
-0.121 & -0.104 & -0.048 & 0.027 & 0.095 & -0.044 & 0.045 & -0.106 & -0.058 & -0.051 \\
-0.031 & -0.031 & -0.031 & 0.051 & 0.037 & -0.023 & 0.062 & -0.030 & -0.011 & 0.023 \\
0.048 & 0.040 & 0.020 & -0.036 & -0.065 & 0.020 & -0.056 & 0.044 & 0.015 & 0.001 \\
-0.031 & -0.028 & -0.016 & 0.010 & 0.025 & -0.016 & 0.020 & -0.030 & -0.022 & -0.028 \\
0.084 & 0.075 & 0.041 & -0.008 & -0.064 & 0.043 & -0.046 & 0.088 & 0.081 & 0.137
\end{array}
\right)
\end{align}
{The Frobenius and $L_1$ norms of this coupling matrix are
$\| C^+_{F} \|_{{F}} \approx 0.4238$ and $\| C^+_{F} \|_{1}  \approx 3.142$, respectively}.
\\

\section{$L_1$-norm-optimized coupling matrices}

The following matrices were obtained as the $L_1$-optimal internetwork coupling matrices $C^+_{1}$ for the tree and random networks. The linear stability was fixed at ${ \Lambda = q = 1 }$.

\subsection{Tree network}

When $c=1$,
\begin{align}
C^+_{1} \approx
\left(
\begin{array}{cccccccccc}
 0.806 & 0.806 & 0.806 & 0.806 & 0.806 & 0.806 & 0.806 & 0.806 & 0.806 & 0.806 \\
 0 & 0 & 0 & 0 & 0 & 0 & 0 & 0 & 0 & 0 \\
 0 & 0 & 0 & 0 & 0 & 0 & 0 & 0 & 0 & 0 \\
 0 & 0 & 0 & 0 & 0 & 0 & 0 & 0 & 0 & 0 \\
 0 & 0 & 0 & 0 & 0 & 0 & 0 & 0 & 0 & 0 \\
 0 & 0 & 0 & 0 & 0 & 0 & 0 & 0 & 0 & 0 \\
 0 & 0 & 0 & 0 & 0 & 0 & 0 & 0 & 0 & 0 \\
 0 & 0 & 0 & 0 & 0 & 0 & 0 & 0 & 0 & 0 \\
 0 & 0 & 0 & 0 & 0 & 0 & 0 & 0 & 0 & 0 \\
 0 & 0 & 0 & 0 & 0 & 0 & 0 & 0 & 0 & 0 \\
\end{array}
\right)
\end{align}
{The $L_1$ and Frobenius norms of this coupling matrix are $\| C^+_{1} \|_1 \approx 8.06$ and $\|C^+_{1}\|_F \approx 2.55$, respectively.}\\

When $c=0.5$,
\begin{align}
C^+_{1} \approx
\left(
\begin{array}{cccccccccc}
 0.500 & 0.500 & 0.500 & 0.500 & 0.500 & 0.500 & 0.500 & 0.500 & 0.500 & 0.500 \\
0.204 &0.204 &0.204 &0.204 &0.204 &0.204 &0.204 &0.204 &0.204 &0.204 \\
0.204 &0.204 &0.204 &0.204 &0.204 &0.204 &0.204 &0.204 &0.204 &0.204 \\
0.204 &0.204 &0.204 &0.204 &0.204 &0.204 &0.204 &0.204 &0.204 &0.204 \\
 0 & 0 & 0 & 0 & 0 & 0 & 0 & 0 & 0 & 0 \\
 0 & 0 & 0 & 0 & 0 & 0 & 0 & 0 & 0 & 0 \\
 0 & 0 & 0 & 0 & 0 & 0 & 0 & 0 & 0 & 0 \\
 0 & 0 & 0 & 0 & 0 & 0 & 0 & 0 & 0 & 0 \\
 0 & 0 & 0 & 0 & 0 & 0 & 0 & 0 & 0 & 0 \\
 0 & 0 & 0 & 0 & 0 & 0 & 0 & 0 & 0 & 0 \\
\end{array}
\right)
\end{align}
{The $L_1$ and Frobenius norms of this coupling matrix are $\| C^+_{1} \|_1 \approx 11.1$ and $\|C^+_{1} \|_F \approx 1.94$, respectively.}\\

When $c=0.25$,
\begin{align}
C^+_{1} \approx \left(
\begin{array}{cccccccccc}
 0.250 & 0.250 & 0.250 & 0.250 & 0.250 & 0.250 & 0.250 & 0.250 & 0.250 & 0.250 \\
 0.250 & 0.250 & 0.250 & 0.250 & 0.250 & 0.250 & 0.250 & 0.250 & 0.250 & 0.250 \\
 0.250 & 0.250 & 0.250 & 0.250 & 0.250 & 0.250 & 0.250 & 0.250 & 0.250 & 0.250 \\
 0.250 & 0.250 & 0.250 & 0.250 & 0.250 & 0.250 & 0.250 & 0.250 & 0.250 & 0.250 \\
 0.121& 0.121& 0.121& 0.121& 0.121& 0.121& 0.121& 0.121& 0.121& 0.121\\
 0.121& 0.121& 0.121& 0.121& 0.121& 0.121& 0.121& 0.121& 0.121& 0.121\\
 0.121& 0.121& 0.121& 0.121& 0.121& 0.121& 0.121& 0.121& 0.121& 0.121\\
 0.121& 0.121& 0.121& 0.121& 0.121& 0.121& 0.121& 0.121& 0.121& 0.121\\
 0.121& 0.121& 0.121& 0.121& 0.121& 0.121& 0.121& 0.121& 0.121& 0.121\\
 0.121& 0.121& 0.121& 0.121& 0.121& 0.121& 0.121& 0.121& 0.121& 0.121\\
\end{array}
\right)
\end{align}
{The $L_1$ and Frobenius norms of this coupling matrix are $\| C^+_{1} \|_1 \approx 17.2$ and $\|C^+_{1} \|_F \approx 1.84$, respectively.}\\

\subsection{Random network}

When $c=1.5$,
\begin{align}
C^+_{1} \approx
\left(
\begin{array}{cccccccccc}
 0 & 0 & 0 & 0 & 0 & 0 & 0 & 0 & 0 & 0 \\
 0 & 0 & 0 & 0 & 0 & 0 & 0 & 0 & 0 & 0 \\
 0 & 0 & 0 & 0 & 0 & 0 & 0 & 0 & 0 & 0 \\
 0 & 0 & 0 & 0 & 0 & 0 & 0 & 0 & 0 & 0 \\
 0 & 0 & 0 & 0 & 0 & 0 & 0 & 0 & 0 & 0 \\
 0 & 0 & 0 & 0 & 0 & 0 & 0 & 0 & 0 & 0 \\
 0 & 0 & 0 & 0 & 0 & 0 & 0 & 0 & 0 & 0 \\
 0 & 0 & 0 & 0 & 0 & 0 & 0 & 0 & 0 & 0 \\
 0 & 0 & 0 & 0 & 0 & 0 & 0 & 0 & 0 & 0 \\
 0 & 0 & 0 & 0 & 0 & 0 & 0 & 0 & 0 & 1.31 \\
\end{array}
\right)
\end{align}
{The $L_1$ and Frobenius norms of this coupling matrix are 
$\| C^+_{1} \|_1 \approx 1.31$ and $\| C^+_{1} \|_F \approx 1.31$, respectively.}\\

When $c=1.0$,
\begin{align}
C^+_{1} \approx
\left(
\begin{array}{cccccccccc}
 0 & 0 & 0 & 0 & 0 & 0 & 0 & 0 & 0 & 0 \\
 0 & 0 & 0 & 0 & 0 & 0 & 0 & 0 & 0 & 0 \\
 0 & 0 & 0 & 0 & 0 & 0 & 0 & 0 & 0 & 0 \\
 0 & 0 & 0 & 0 & 0 & 0 & 0 & 0 & 0 & 0 \\
 0 & 0 & 0 & 0 & 0 & 0 & 0 & 0 & 0 & 0 \\
 -0.35 & 0 & 0 & 0 & 0 & 0 & 0 & 0 & 0 & 0 \\
 0 & 0 & 0 & 0 & 0 & 0 & 0 & 0 & 0 & 0 \\
 0 & 0 & 0 & 0 & 0 & 0 & 0 & 0 & 0 & 0 \\
 0 & 0 & 0 & 0 & 0 & 0 & 0 & 0 & 0 & 0 \\
 0 & 0 & 0 & 0 & 0 & 0 & 0 & 0 & 0 & 1.0 \\
\end{array}
\right)
\end{align}
{The $L_1$ and Frobenius norms of this coupling matrix are $\| C^+_{1} \|_1 \approx 1.35$ and $\| C^+_{1} \|_F \approx 1.06$, respectively.}\\

When $c=0.5$,
\begin{align}
C^+_{1} \approx
\left(
\begin{array}{cccccccccc}
 0 & 0 & 0 & 0 & 0 & 0 & 0 & 0 & 0 & 0 \\
 0 & 0 & 0 & 0 & 0 & 0 & 0 & 0 & 0 & 0 \\
 0 & 0 & 0 & 0 & 0 & 0 & 0 & 0 & 0 & 0 \\
 0 & 0 & 0 & 0 & 0 & 0 & 0 & 0 & 0 & 0 \\
 0 & 0 & 0 & 0 & 0 & 0 & 0 & 0 & 0 & 0 \\
 -0.50 & 0 & 0 & 0 & 0 & 0 & 0 & -0.48 & 0 & 0 \\
 0 & 0 & 0 & 0 & 0 & 0 & 0 & 0 & 0 & 0 \\
 0 & 0 & 0 & 0 & 0 & 0 & 0 & 0 & 0 & 0 \\
 0 & 0 & 0 & 0 & 0 & 0 & 0 & 0 & 0 & 0 \\
 0 & 0 & 0 & 0 & 0 & 0 & 0 & 0 & 0 & 0.50 \\
\end{array}
\right)
\end{align}
{The $L_1$ and Frobenius norms of this coupling matrix are $\| C^+_{1} \|_1 \approx 1.48$ and $\| C^+_{1} \|_F \approx 0.852$, respectively.}
\\

\section{Comparison of the two types of optimization}

Tables~\ref{normtable1} and~\ref{normtable2} summarize the Frobenius and $L_1$ norms of the coupling matrices $C^d, C^+_F$, and $C^+_1$ that realize the given linear stability $\Lambda = q = 1$ for the tree and random networks, respectively.
Tables~\ref{normtable1b} and~\ref{normtable2b} summarize the linear stability $\Lambda$ that can be attained by using the normalized coupling matrices $C^d / \| C^d \|$, $C^+_F / | \| C^+_F \|$, and $C^+_1 / | \| C^+_1 \|$, where the norm $\| \cdot \|$ is either Frobenius or $L_1$.
The Frobenius-norm-optimization provides the best results when evaluated by the Frobenius norm and the $L_1$-norm-optimization gives the best results when evaluated by the $L_1$ norm.
It is noted that the $L_1$-norm optimization with weight constraints provides reasonably good coupling matrices also in terms of the Frobenius norm.
\begin{table}[h]
\centering
\begin{tabular}[t]{lcc}
\hline
& Frobenius norm \quad & \quad $L_1$ norm\\
\hline
No optimization (diagonal) $C^d$ &6.37 & 20.2 \\
Frobenius-norm optimization $C^+_F$ &{\bf 1.75} &15.2\\
$L_1$-norm optimization $C^+_1$ ($c=1.0$) &2.55 & {\bf 8.06} \\
$L_1$-norm optimization $C^+_1$ ($c=0.5$) &1.94 &11.1\\
$L_1$-norm optimization $C^+_1$ ($c=0.25$) & 1.84 &17.2\\
\hline
\end{tabular}
\caption{Frobenius and $L_1$ norms of the coupling matrices for the tree network.}
\label{normtable1}
\end{table}
\begin{table}[h]
\centering
\begin{tabular}[t]{lcc}
\hline
& $\Lambda$ (Frobenius normalized) \quad & \quad $\Lambda$ ($L_1$ normalized) \\
\hline
No optimization (diagonal) $C^d/\|C^d\|$ &0.157 & 0.0496 \\
Frobenius-norm optimization $C^+_F/\|C^+_F\|$ & {\bf 0.570} &0.0659\\
$L_1$-norm optimization $C^+_1/\|C^+_1\|$ ($c=1.0$) &0.392 &{\bf 0.124} \\
$L_1$-norm optimization $C^+_1/\|C^+_1\|$ ($c=0.5$) &0.515 &0.0901\\
$L_1$-norm optimization $C^+_1/\|C^+_1\|$ ($c=0.25$) & 0.543 &0.0581\\
\hline
\end{tabular}
\caption{Linear stability $\Lambda$ of the normalized coupling matrices for the tree network.}
\label{normtable1b}
\end{table}

\begin{table}[h]
\centering
\begin{tabular}[t]{lcc}
\hline
& Frobenius norm \quad & \quad $L_1$ norm\\
\hline
No optimization (diagonal) $C^d$ &3.59 & 11.3 \\
Frobenius-norm optimization $C^+_F$ &{\bf 0.424} &3.14 \\
$L_1$-norm optimization $C^+_1$ ($c=1.5$) &1.31 &{\bf 1.31}\\
$L_1$-norm optimization $C^+_1$ ($c=1.0$) &1.06 &1.35\\
$L_1$-norm optimization $C^+_1$ ($c=0.5$) &0.852 &1.48\\
\hline
\end{tabular}
\caption{Frobenius and $L_1$ norms of the coupling matrices for the random network.}
\label{normtable2}
\end{table}
\begin{table}[b]
\centering
\begin{tabular}[t]{lcc}
\hline
& $\Lambda$ (Frobenius normalized) \quad & \quad $\Lambda$ ($L_1$ normalized) \\
\hline
No optimization (diagonal) $C^d/\|C^d\|$ &0.278 & 0.0881  \\
Frobenius-norm optimization $C^+_F/\|C^+_F\|$ & {\bf 2.36 } &0.318\\
$L_1$-norm optimization $C^+_1/\|C^+_1\|$ ($c=1.0$) &0.763 &{\bf 0.763} \\
$L_1$-norm optimization $C^+_1/\|C^+_1\|$ ($c=0.5$) &0.943 &0.740\\
$L_1$-norm optimization $C^+_1/\|C^+_1\|$ ($c=0.25$) & 1.17  &0.676\\
\hline
\end{tabular}
\caption{Linear stability $\Lambda$ of the normalized coupling matrices for the tree network.}
\label{normtable2b}
\end{table}

\clearpage


%

\end{document}